\documentclass[11pt]{article}
\pdfoutput=1

\usepackage[letterpaper,margin=1in,bottom=1.4in]{geometry} 
\usepackage[format=plain,font=small,labelfont=bf,textfont=up]{caption}

\usepackage[utf8]{inputenc} 
\usepackage[T1]{fontenc}    
\usepackage[hidelinks]{hyperref}       
\usepackage{url}            
\usepackage{booktabs}       
\usepackage{amsfonts}       
\usepackage{nicefrac}       
\usepackage{microtype}      
\usepackage[affil-it]{authblk} 
\usepackage{tikz}
\usepackage{lmodern}
\usepackage{float}

\usepackage[linesnumbered,ruled,vlined]{algorithm2e}

\SetCommentSty{mycommfont}

\usepackage{amsmath}
\usepackage{amssymb}
\usepackage{amsthm}
\usepackage{mathtools}
\mathtoolsset{showonlyrefs,showmanualtags}
\allowdisplaybreaks
\usepackage{bbm}
\usepackage{graphicx}
\usepackage{subcaption}
\usepackage{thmtools, thm-restate}  
\usepackage[toc,page,header]{appendix}
\usepackage{minitoc}

\newcommand{\texpath}{tex}
\newcommand{\figpath}{figs}

\newcounter{spacesave}
\usepackage[style=authoryear,sorting=nyt,sortcites=false, natbib=true, backend=bibtex, maxbibnames=99]{biblatex}
\DeclarePairedDelimiter\paren\lparen\rparen

\DeclarePairedDelimiter\braces\lbrace\rbrace

\DeclarePairedDelimiter\abs\lvert\rvert

\providecommand{\bbone}{\mathbf{1}}
\DeclarePairedDelimiterXPP\indicator[1]{\bbone}{\lbrack}{\rbrack}{}{#1}

\DeclarePairedDelimiterXPP\expf[1]{\exp}{\lparen}{\rparen}{}{#1}
\DeclarePairedDelimiterXPP\logf[1]{\log}{\lparen}{\rparen}{}{#1}
\DeclarePairedDelimiterXPP\maxf[1]{\max}{\lparen}{\rparen}{}{#1}
\DeclarePairedDelimiterXPP\minf[1]{\min}{\lparen}{\rparen}{}{#1}

\DeclareMathOperator*{\sgn}{sgn}
\DeclarePairedDelimiterXPP\sgnf[1]{\sgn}{\lparen}{\rparen}{}{#1}

\DeclarePairedDelimiterXPP\func[2]{#1}{\lparen}{\rparen}{}{#2}

\DeclareMathOperator*{\atan}{atan}
\DeclarePairedDelimiterXPP\atanf[1]{\atan}{\lparen}{\rparen}{}{#1}
\DeclarePairedDelimiterXPP\tanf[1]{\tan}{\lparen}{\rparen}{}{#1}

\DeclarePairedDelimiter\setb\lbrace\rbrace

\newcommand{\Reals}{\mathbb{R}}

\DeclareMathOperator*{\argmin}{arg\,min}

\makeatletter
\newcommand*{\tran}{{\mathpalette\@tran{}}}
\newcommand*{\@tran}[2]{\raisebox{\depth}{$\m@th#1\intercal$}}
\makeatother

\DeclarePairedDelimiter\norm\lVert\rVert
\DeclarePairedDelimiterXPP\tnorm[1]{}{\lVert}{\rVert_{1}}{}{#1}
\DeclarePairedDelimiterXPP\enorm[1]{}{\lVert}{\rVert_{2}}{}{#1}
\DeclarePairedDelimiterXPP\inorm[1]{}{\lVert}{\rVert_{\infty}}{}{#1}
\DeclarePairedDelimiterXPP\pnorm[2]{}{\lVert}{\rVert_{#1}}{}{#2}
\DeclarePairedDelimiterXPP\opnorm[1]{}{\lVert}{\rVert_{op}}{}{#1}

\DeclarePairedDelimiterXPP\detf[1]{\det}{\lparen}{\rparen}{}{#1}

\DeclarePairedDelimiterXPP\kerf[1]{\ker}{\lparen}{\rparen}{}{#1}

\DeclareMathOperator{\trsym}{tr}
\DeclarePairedDelimiterXPP\tr[1]{\trsym}{\lparen}{\rparen}{}{#1}

\DeclareMathOperator{\vspansym}{span}
\DeclarePairedDelimiterXPP\vspan[1]{\vspansym}{\lparen}{\rparen}{}{#1}

\DeclareMathOperator{\diagsym}{diag}
\DeclarePairedDelimiterXPP\diag[1]{\diagsym}{\lparen}{\rparen}{}{#1}

\DeclareMathOperator{\ranksym}{rank}
\DeclarePairedDelimiterXPP\rank[1]{\ranksym}{\lparen}{\rparen}{}{#1}

\DeclareMathOperator{\vectorizesym}{vec}
\DeclarePairedDelimiterXPP\vectorize[1]{\vectorizesym}{\lparen}{\rparen}{}{#1}

\DeclareMathOperator*{\esssup}{ess\,sup}
\DeclarePairedDelimiterXPP\esssupf[1]{\esssup}{\lparen}{\rparen}{}{#1}

\let\Prsym\Pr
\let\Pr\relax
\DeclarePairedDelimiterXPP\Pr[1]{\Prsym}{\lparen}{\rparen}{}{%
	#1}

\DeclarePairedDelimiterXPP\Prsub[2]{\Prsym_{#1}}{\lparen}{\rparen}{}{%
	#2}

\DeclareMathOperator{\Esym}{E}
\DeclarePairedDelimiterXPP\E[1]{\Esym}{\lbrack}{\rbrack}{}{%
	#1}

\DeclarePairedDelimiterXPP\Esub[2]{\Esym_{#1}}{\lbrack}{\rbrack}{}{%
	#2}

\DeclareMathOperator{\Varsym}{Var}
\DeclarePairedDelimiterXPP\Var[1]{\Varsym}{\lparen}{\rparen}{}{%
	#1}

\DeclarePairedDelimiterXPP\Varsub[2]{\Varsym_{#1}}{\lparen}{\rparen}{}{%
	#2}

\DeclarePairedDelimiterXPP\EstVar[1]{\widehat{\Varsym}}{\lparen}{\rparen}{}{%
	#1}

\DeclareMathOperator{\Covsym}{Cov}
\DeclarePairedDelimiterXPP\Cov[1]{\Covsym}{\lparen}{\rparen}{}{%
	#1}

\DeclarePairedDelimiterXPP\Covsub[2]{\Covsym_{#1}}{\lparen}{\rparen}{}{%
	#2}

\DeclareMathOperator{\Corrsym}{Corr}
\DeclarePairedDelimiterXPP\Corr[1]{\Corrsym}{\lparen}{\rparen}{}{%
	#1}

\makeatletter
\newcommand{\indep}{\protect\mathpalette{\protect\@indep}{\perp}}
\newcommand*{\@indep}[2]{\mathrel{\rlap{$#1#2$}\mkern3mu{#1#2}}}
\makeatother

\newcommand{\darrow}{\overset{d}{\longrightarrow}}
\newcommand{\parrow}{\overset{p}{\longrightarrow}}

\newcommand{\bigOsym}{\mathcal{O}}
\DeclarePairedDelimiterXPP\bigO[1]{\bigOsym}{\lparen}{\rparen}{}{#1}
\DeclarePairedDelimiterXPP\bigOt[1]{\widetilde{\bigOsym}}{\lparen}{\rparen}{}{#1}

\newcommand{\littleOsym}{o}
\DeclarePairedDelimiterXPP\littleO[1]{\littleOsym}{\lparen}{\rparen}{}{#1}

\newcommand{\bigOpsym}{\bigOsym_p}
\DeclarePairedDelimiterXPP\bigOp[1]{\bigOpsym}{\lparen}{\rparen}{}{#1}

\newcommand{\littleOpsym}{\littleOsym_p}
\DeclarePairedDelimiterXPP\littleOp[1]{\littleOpsym}{\lparen}{\rparen}{}{#1}

\newcommand{\bigOmegasym}{\Omega}
\DeclarePairedDelimiterXPP\bigOmega[1]{\bigOmegasym}{\lparen}{\rparen}{}{#1}
\DeclarePairedDelimiterXPP\bigOmegap[1]{\bigOmegasym_p}{\lparen}{\rparen}{}{#1}

\newcommand{\littleOmegasym}{\omega}
\DeclarePairedDelimiterXPP\littleOmega[1]{\littleOmegasym}{\lparen}{\rparen}{}{#1}

\newcommand{\bigThetasym}{\Theta}
\DeclarePairedDelimiterXPP\bigTheta[1]{\bigThetasym}{\lparen}{\rparen}{}{#1}

\DeclarePairedDelimiterXPP\cosf[1]{\cos}{\lparen}{\rparen}{}{#1}
\DeclarePairedDelimiterXPP\sinf[1]{\sin}{\lparen}{\rparen}{}{#1}

\newcommand{\quadtext}[1]{\quad\text{#1}\quad}

\newcommand{\quadand}{\quadtext{and}}

\theoremstyle{plain}
\newtheorem{theorem}{Theorem}[section]

\newtheorem{corollary}[theorem]{Corollary}
\newtheorem{lemma}[theorem]{Lemma}
\newtheorem{proposition}[theorem]{Proposition}

\newtheorem{innerrefproposition}{Proposition}
\newenvironment{refproposition}[1]
{\renewcommand\theinnerrefproposition{#1}\innerrefproposition}
{\endinnerrefproposition}

\newtheorem{innerreftheorem}{Theorem}
\newenvironment{reftheorem}[1]
{\renewcommand\theinnerreftheorem{#1}\innerreftheorem}
{\endinnerreftheorem}

\theoremstyle{definition}
\newtheorem{assumption}{Assumption}
\newtheorem*{assumption*}{Assumption}

\theoremstyle{remark}

 \newcommand{\Tpre}{T_{\mathrm{pre}}}
 \newcommand{\Tpost}{T_{\mathrm{post}}}
 
 \newcommand{\popdist}{\mathcal{P}}

 \newcommand{\obsout}[2]{Y_{#1}^{(#2)}}

 \newcommand{\poout}[3]{ \obsout{#1}{#2}(#3) }
 
 \newcommand{\estimandsym}{\tau}
 \newcommand{\att}[1]{\estimandsym^{(#1)}_{\mathrm{ATT}}}
 \newcommand{\attt}{\att{t}}
 \newcommand{\totalatt}{\estimandsym_{\mathrm{ATT}}}
  \newcommand{\linatt}{\estimandsym^{\mathrm{Lin}}_{\mathrm{ATT}}}

 \newcommand{\dd}[1]{ \estimandsym^{(#1)}_{\mathrm{DD}} }
 \newcommand{\ddt}{ \dd{t} }
 \newcommand{\estdd}[1]{\widehat{\estimandsym}^{(#1)}_{\mathrm{DD}}}
 \newcommand{\estddt}{\estdd{t}}
 \newcommand{\totaldd}{ \estimandsym_{\mathrm{DD}}}
 \newcommand{\esttotaldd}{ \widehat{\estimandsym}_{\mathrm{DD}} }
\newcommand{\lindd}{\estimandsym^{\mathrm{Lin}}_{\mathrm{DD}}}

  \newcommand{ \overvio }[1]{ \Delta^{(#1)} }
  \newcommand{\overviot}{\overvio{t}}
  \newcommand{ \itervio }[1]{ R^{(#1)} }
  \newcommand{\iterviot}{\itervio{t}} 
  \newcommand{\titervio}[1]{ \tilde{R}^{(#1)} }

  \newcommand{\estitervio}[1]{\widehat{R}^{(#1)}}
  \newcommand{\estiterviot}{\estitervio{t}}
  
  \newcommand{\estovervio}[1]{\widehat{\Delta}^{(#1)}}

  \newcommand{\popfunc}{\theta}
  \newcommand{\estpopfunc}{\widehat{\theta}}
    \newcommand{\popfuncn}{\theta_n}
  \newcommand{\estpopfuncn}{\widehat{\theta}_n}
  
  \newcommand{\pfcov}{\Sigma_\popfunc}
  
  \newcommand{\estcov}{\widehat{\Sigma}}
  \newcommand{\covsig}{\Sigma}
  
 \newcommand{\Spre}{S_{\mathrm{pre}}}
 \newcommand{\Spost}{S_{\mathrm{post}}}
 \newcommand{\estSpre}{\widehat{S}_{\mathrm{pre}}}
  
 \newcommand{\extrapbias}{\kappa}
  \newcommand{\extrapbiaslin}{\kappa^{\mathrm{Lin}}}

 \newcommand{\crv}{f(\alpha, \covsig)}
  \newcommand{\hcrvn}{f(\alpha, \estcov)}

\newcommand{\level}{\alpha}
 \newcommand{\ci}{\widehat{C}_{\level}}

\usepackage{expcmd}
\newcommand\mainref\ref
\newcommand\suppref\ref


\title{Valid Inference when Testing Violations of Parallel Trends for Difference-in-Differences}
\author[1]{Jonas M. Mikhaeil}
\author[1]{Christopher Harshaw}
\affil[1]{Columbia University}
\date{\today}

\usepackage{xcolor}         

\begin{document}
	
	\makeatletter%
	
	\begin{NoHyper}\gdef\@thefnmark{}\@footnotetext{\hspace{-1em}We thank
Tim Armstrong,
Brantly Callaway,
Cl{\'e}ment de Chaisemartin,
Donald Green,
Hyunseung Kang,
Jonathan Roth,
Pedro Sant'Anna,
and
Fredrik S{\"a}vje
for insightful comments and stimulating discussions which helped shape this work.
Christopher Harshaw gratefully acknowledges funding from NSF grant MMS-2316335.}\end{NoHyper}%
	\makeatother%
	
	\maketitle
	\thispagestyle{empty}
	
	\begin{abstract}
		The difference-in-differences (DID) research design is a key identification strategy which allows researchers to estimate causal effects under the parallel trends assumption.
While the parallel trends assumption is counterfactual and cannot be tested directly, researchers often examine pre-treatment periods to check whether the time trends are parallel before treatment is administered.
A recent literature has shown that existing preliminary tests have adverse effects on conventional statistical methods for estimation and inference, including low power, bias, and undercoverage.
In this paper, we describe simple preliminary tests and corresponding confidence intervals for the causal effect which overcome these issues.
Under mild separation conditions, the preliminary test is shown to be consistent and the confidence intervals for the causal effect have valid coverage conditional on passing the test.
Our results hold under what we refer to as the conditional extrapolation assumption, which posits a relationship between the unidentified post-treatment violation of parallel trends and the identified pre-treatment violations. 
We view the conditional extrapolation assumption as one formalization of the assumption which is implicitly held when conducting a preliminary test for parallel trends.
To illustrate the performance of the proposed methods, we use synthetic data as well as data on recentralization of public services in Vietnam and right-to-carry laws in Virginia.

	\end{abstract}

	\newpage
	
	\pagenumbering{roman}
	
	\doparttoc 
	\faketableofcontents 
	\part{} 
	\parttoc 
	
	\newpage

	\pagenumbering{arabic}
	
	\section{Introduction}

The difference-in-differences (DID) research design is a widely popular identification strategy that allows researchers to estimate causal effects in the presence of unmeasured confounding, i.e. selection on unobservables.
This identification strategy, however, crucially relies on a \textit{parallel trends} assumption: the counterfactual time-trend in the treatment group had it not received treatment is assumed to be the same as the time-trend in the control group. 

While this parallel trends assumption is counterfactual and cannot be tested directly, researchers often have access to multiple pre-treatment time-periods. When this is the case, it is common practice to test whether trends in the treatment- and control group are parallel in the periods leading up to treatment \citep{angrist_mostly_2009,gallagher_learning_2014, markevich_economic_2018,egami_using_2023}.
If the preliminary tests yield evidence of violations of the parallel trends in the pre-treatment period, researchers may doubt the plausibility of the parallel trends assumption in the post-treatment period.
On the other hand, preliminary tests which do not yield evidence of pre-treatment parallel trends violations may give credence to the parallel trends assumption in the post-treatment period.

However, a recent literature has cautioned researchers against the use of preliminary tests for violations of parallel trends in pre-treatment periods. 
Recent literature has shown that conventional preliminary tests will have adverse effects on conventional statistical methods for estimation and inference, e.g. low power, bias, and severe undercoverage \citep{freyaldenhoven_pre-event_2019,bilinski_nothing_2020,kahn-lang_promise_2020,roth_pretest_2022}.
This presents an issue for applied researchers: while preliminary tests are intuitive from an empirical perspective, conventional methods suffer from statistical deficiencies which make them unattractive.

In this paper, we propose a preliminary test and confidence intervals for the causal effect which do not suffer from these statistical shortcomings.
The analyst first performs a preliminary test based on the magnitude of the pre-treatment violations to determine whether further DID analysis is justified.
The preliminary test is both conceptually simple and easy to implement.
We show that the test is asymptotically consistent, which is to say that both Type I and Type II errors vanish on suitably separated null and alternative hypotheses.
If the preliminary test passes, then the analyst reports the proposed confidence intervals, whose widths take into account the estimated magnitude of the pre-treatment violations.
We show that the confidence intervals are asymptotically valid after conditioning on passing the preliminary test.
This resolves one of the issues with conventional tests and intervals identified by \citet{freyaldenhoven_pre-event_2019,roth_pretest_2022}.
The width of the confidence interval depends on the sample size as well as the magnitude of the pre-treatment violations of parallel trends.
In the case that pre-treatment violations of parallel trends are negligible, the width of these intervals converges to zero in large samples at the usual $n^{-1/2}$ rate.

Our results hold under what we refer to as the \emph{conditional extrapolation assumption}.
The conditional extrapolation assumption posits a relationship between the unidentified post-treatment violations of parallel trends and the identified pre-treatment violations.
An informal version of the assumption, formalized in Section \ref{Sec:ExtrapolationAssumption}, is given below:
\begin{assumption*}[Conditional Extrapolation, Informal]
	If the pre-treatment violations of parallel trends are \emph{below a pre-specified acceptable level}, then post-treatment violations are not more severe than the pre-treatment violations.
\end{assumption*}
The \emph{extrapolation condition} is the condition that the pre-treatment violations are below the acceptable level.
The most salient feature of the conditional extrapolation assumption is that if the extrapolation condition does not hold, then nothing can be said about the post-treatment violations, i.e. they may be arbitrarily large.
In this case, DID estimators can have arbitrarily large bias and thus DID is deemed inappropriate as a research design.
One may view the conditional extrapolation assumption through the lens of falsification: the analyst has a justification for why parallel trends ought to hold and sufficiently large violations in the pre-treatment period falsify this justification.

To the best of our knowledge, the conditional extrapolation assumption presented in this paper is the first formal assumption in the literature that necessitates preliminary testing.
In particular, the analyst must run a preliminary test to determine whether the extrapolation condition{\textemdash}namely, that the pre-treatment violations are below the acceptable level{\textemdash}is indeed satisfied before carrying out further DID analysis.
In this sense, one may interpret our conditional extrapolation assumption as one possible formalization of the assumption which is implicitly made when applied researchers perform a preliminary test.
From this perspective, the contribution of the paper may be understood as providing a formalization of this implicitly held assumption and providing appropriate testing and inference procedures.

It is instructive to compare our conditional extrapolation assumption with two assumptions made in the literature. 
Under the classical parallel trends assumption, there are no post-treatment violations of parallel trends by assumption; any investigation of pre-treatment trends is irrelevant.
There are also partial identification-type assumptions under which the post-treatment violations are never more severe than the pre-treatment violations \citep{manski_how_2018, rambachan_more_2023}.
Under this assumption, there is no need for preliminary testing of pre-treatment parallel trends violations -- they are assumed to always extrapolate to the post-treatment period.
If one takes as a methodological given that pre-treatment violations ought to be evaluated via some sort of preliminary test because extrapolation from pre-treatment violations to post-treatment violations is not always substantively justified, then neither of these assumptions offers the right prescription.

\subsection{Illustration} \label{sec:illustration}

\begin{figure}[tbh]
    \centering
    \includegraphics[width=\linewidth]{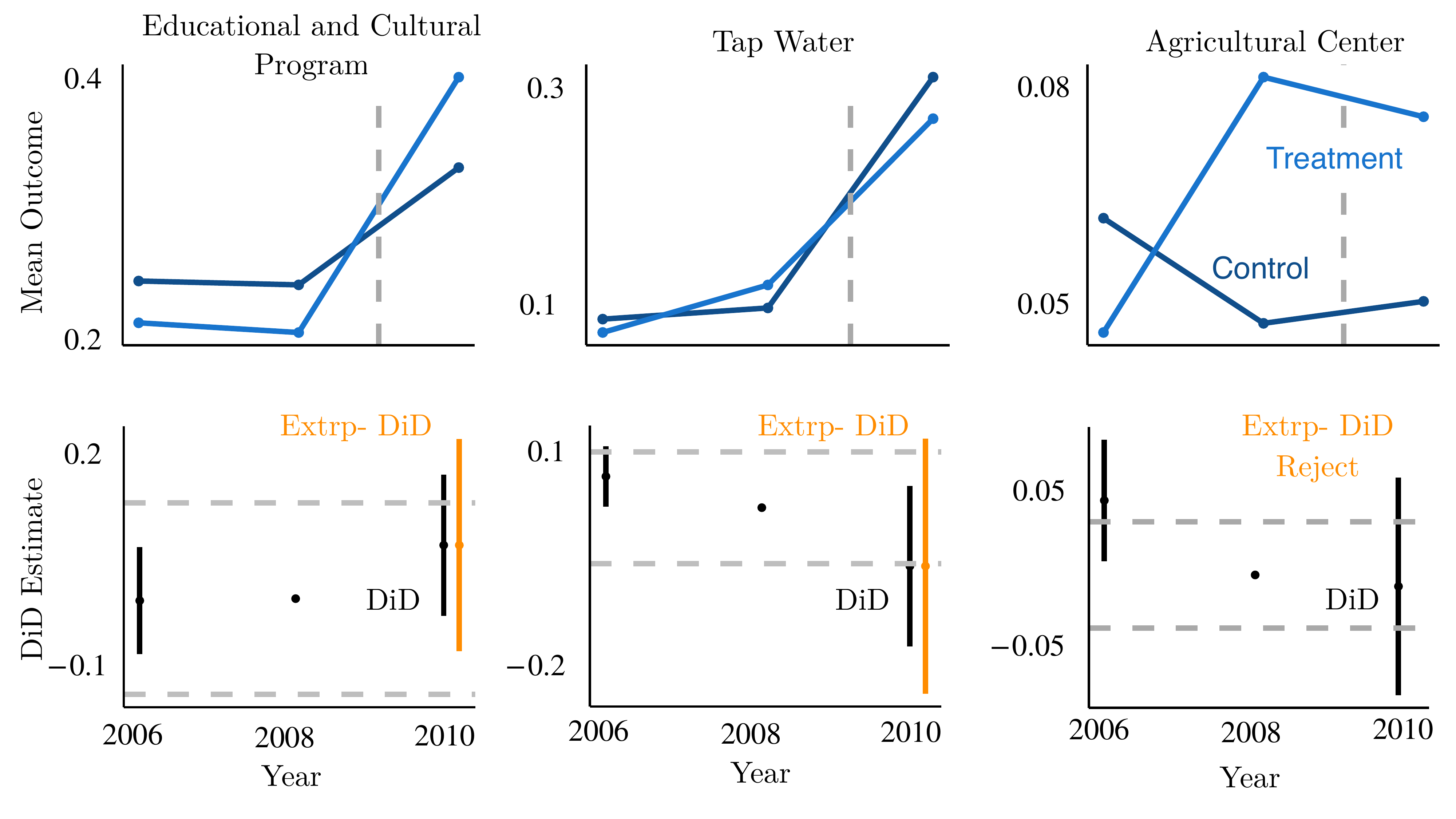}
    \caption{
    	Upper panel: evolution of the means of three outcomes studied by \citet{Malesky_Nguyen_Tran_2014}.
    	Lower Panel: comparison of classical DID inference (black) with our $95$\% confidence intervals based on the extrapolation assumption (orange). For the agricultural center outcome our preliminary test rejects the extrapolation condition, and we do not report any inference based on the conditional extrapolation assumption.
    	Gray, dashed lines indicate the threshold for acceptable violations of parallel trends in the pre-treatment period.}
    \label{fig:malesky}
\end{figure}

To illustrate our approach, we re-analyze the effect of recentralization on public services in Vietnam \citep{Malesky_Nguyen_Tran_2014}. In the late 2000s, Vietnam's central government proposed the dissolution of District People's Councils (DPCs) -- a form of local governance -- to improve economic development.
This was highly contentious and, in 2009, Vietnamese leadership officiated a pilot program to study the effects of recentralization.
Officials stratified the selection of provinces that would be recentralized based on their region and other covariates deemed important.
This careful selection of treated (that is, recentralized) provinces lends some ex ante plausibility to the parallel trends assumption.
Even with carefully chosen treatment rollout, it is likely that some outcomes of interest may not have evolved in parallel absent of treatment, so that investigations into pre-treatment trends are warranted.

For the purposes of our illustration, we focus on three outcomes of interest: existence of educational and cultural programs, whether tap water is the main source of drinking water, and existence of an agricultural extension center.
The top panels in Figure~\ref{fig:malesky} show how each of these three observed outcomes in the treatment and control group change over time, with a dashed line representing the treatment time.
Visually speaking, it appears that the pre-treatment trends are almost exactly parallel in the first outcome, slightly deviating from parallel in the second outcome, and severely non-parallel in the third outcome.

We begin by conducting the proposed preliminary test to determine whether the extrapolation condition holds, i.e. that the severity of the pre-treatment parallel trend violation falls below a pre-determined threshold.
If the test passes, then extrapolation of the severity of parallel-trend violations to the post-treatment period is considered to be justified and we move on to constructing confidence intervals for the Average Treatment Effect on the Treated (ATT) in this period.
These intervals are centered at the difference-in-differences estimate and their widths depend not only on the estimated severity of parallel trend violations, but also the statistical uncertainty in estimating these quantities.
We refer to Appendix~\suppref{app:empirical-example} for more details on the implementation of our procedures in this illustration.

The bottom panel of Figure~\ref{fig:malesky} includes the severity test, the resulting confidence interval (when the test passes) and a conventional confidence interval for comparison.
The dot appearing at 2006 is the estimate of the severity of the parallel trends violation and the bars depict the estimated standard deviation of the estimate.
The dashed gray line represents the acceptable violation of parallel trends that warrants extrapolation, which was chosen to be half of the outcome.
The conventional confidence intervals are the black lines appearing in 2010 and our proposed confidence intervals are the orange lines appearing in 2010.

For the first two outcomes, the preliminary test declares the extrapolation condition to be true, so that extrapolation is deemed to be warranted and we proceed to construct the confidence intervals.
Our confidence interval is larger than the conventional interval, because the latter does not take into account the pre-treatment violations nor their statistical uncertainty.
In the first example, the violation is estimated to be close to zero but the statistical uncertainty in this estimate is large.
In the second example, the violation is estimated to be larger, but there is less statistical uncertainty.
Both confidence intervals in this illustration contain 0, which supports the hypothesis that recentralization had little effect on educational and cultural programs or on tap water.
For the third outcome, the pre-treatment violation of parallel trends is so great that we do not proceed to make any inference on the effect of recentralization on agricultural centers. 
In this illustration, the results of the preliminary test align with the visual inspection of parallel trends.

\subsection{Related Work}
The use of difference-in-differences (DID) as a research design has a long history \citep[see e.g., ][]{Snow_1855, Obenauer_Nienburg, Ashenfelter_85, Card_90}.
When pre-treatment periods are available, it is typical to check for violations of parallel trends in the pre-treatment period \citep{angrist_mostly_2009,Card_Krueger_2000}. In particular, many empirical studies report event-study plots to lend plausibility to the parallel trends assumption \citep{Autor_03,fitzpatrick_early_2014,gallagher_learning_2014,markevich_economic_2018}.

An emerging research literature has cautioned against the use of pre-tests for violations of parallel trends in the pre-treatment period altogether  as they may have adverse effects on classical DID inference, such as bias and severe undercoverage \citep{bilinski_nothing_2020,kahn-lang_promise_2020,freyaldenhoven_pre-event_2019,roth_pretest_2022}. 
For a further discussion, we recommend the review of \citet{Roth_2023b} and references therein.
Beyond the difference-in-differences method, preliminary testing of modeling and identifying assumptions has a long history in econometrics \citep[see e.g.,][]{Giles_93}.
For example, there have been various investigations into statistical inference after preliminary testing in an instrumental variable analysis \citep{guggenberger_impact_2010, Andrews_18, Bi_2020}.
\citet{Armstrong_2025} propose an alternative to preliminary testing by constructing estimators that adapt to misspeciﬁcation.

In response to this literature, \citet{bilinski_nothing_2020} and \citet{Holger_2020} develop equivalence, or non-inferiority, tests that reverse the burden of proof when assessing parallel trends: rather than failing to reject equality of trends, researchers must provide evidence that deviations are substantively negligible.
In contrast, our approach does not aim to validate the parallel trends assumption.
Instead, we replace the exact parallel trends assumption with a conditional extrapolation assumption.
In our approach, the goal of preliminary testing is to determine whether pre-treatment violations are acceptable for extrapolation to the post-treatment period, and our confidence intervals provide valid coverage conditional on passing this test.

\citet{Chaisemartin_26} show that, under an exact null of valid specification, inference conditional on not rejecting a broad class of pre-tests remains valid, albeit possibly conservative. Their results do not immediately apply to our setting because our extrapolation condition defines a composite inequality null, rather than an exact null of valid specification. Whether a modified version of their proof strategy can yield conditional coverage guarantees for our setting remains an open question.

Instead of testing for violations of parallel trends, authors have proposed alternative methods of inference or identification. 
One approach is to pursue partial identification \citep{manski_how_2018,rambachan_more_2023}.
Our approach is most similar to the work of \citet{rambachan_more_2023}, who develop inference procedures for difference-in-differences designs under restrictions linking post-treatment violations of parallel trends to pre-treatment differences. 
Our work differs in its inferential goal: rather than ensuring worst-case \emph{unconditional} coverage over all data-generating processes satisfying specified restrictions on the joint pre- and post-treatment violations, we provide \emph{conditional} coverage guarantees given that the observed pre-treatment violations satisfy an extrapolation condition. In particular, our procedure adapts inference to the realized pre-treatment estimates via a preliminary test, whereas their approach remains uniform over all admissible violations and does not condition its coverage guarantees on the observed pre-treatment realization. Additionally, while their moment-inequality procedures rely on polyhedral restrictions, our conditional extrapolation assumption allows for non-polyhedral restrictions not directly handled by their moment-inequality approach.

Similarly, our approach is related to the literature on bias-aware confidence intervals \citep{Donoho_94,Armstrong_18,Armstrong_20}. A key difference is that in our setting the worst-case bias bound is not treated as a fixed quantity, but is itself estimated and must be handled inside the coverage constraint jointly with the estimator.

More broadly, our work contributes to a growing literature that seeks to assess, relax, or theoretically justify the parallel trends assumption when its exact validity is uncertain.
\citet{keele2019} perform a sensitivity analysis quantifying how much bias from violations of parallel trends is necessary to change the conclusion of a study.
\citet{Roth_2023} develop a test for the null hypothesis that parallel trends are insensitive to functional form.
\citet{Kwon_Roth_2024} suggest an empirical Bayes approach in which researchers update their beliefs about the post-treatment violations of parallel trends given the observed pre-treatment violations.
\citet{ham2024benefitscostsmatchingprior} study the effects of matching when parallel trends are violated.
\citet{freyaldenhoven_pre-event_2019} circumvent reliance on the parallel trends assumption altogether by making use of a covariate that is related to the confounder but unaffected by treatment.
\citet{Marx_24, Ghanem_2025} take a theoretical approach to evaluating the parallel trends assumption by investigating which economic models of selection and choice are compatible with the assumption.

In this paper, we consider event studies with a block-adoption design where treatment occurs for all subjects at one time.
However, there is a growing current literature which investigates staggered-adoption designs where treatment rollout is not simultaneous but, instead, occurs in a staggered manner.
Classical DID approaches fail to recover the relevant treatment estimands and alternative estimators have been proposed \citep{Sun_Abraham_2021,Goodman-Bacon_2021,Callaway_SantAnna_2021,Athey_Imbens_2022,Borusyak_Jaravel_Spiess_2024}.
While the parallel trends assumption is untestable in the block-adoption design, some formulations of parallel trends in the staggered setting imply overidentifying restrictions that can be partially tested \citep{marcus_2020,chen2025}.

When DID is determined not to be a valid research design, there are additional methodologies which may be appropriate, including synthetic control methods \citep{Abadie_2010,Abadie_2021} and Bayesian panel-data models \citep{BenMichael_2023,Bell_2025}.

	\section{Preliminaries}
\label{sec:prelim}
We adopt a potential outcomes framework for the difference-in-differences (DID) design \citep{abadie_semiparametric_2005,egami_using_2023}.
However, all of our results are also applicable under a two-way fixed-effects structural equation model.
We describe this connection in Appendix~\suppref{App:Equiv}.

\subsection{Causal Estimands in Repeated Cross-Sectional and Panel Data} \label{sec:causal-estimands-and-setup}

We assume that the analyst is working with $T$ time periods indexed as $t = 1 \dots T$.
We use $t_0$ to denote the time at which the treatment was administered.
The first $\Tpre = t_0 - 1$ observation times are referred to as the \emph{pre-treatment} period and the remaining $\Tpost = T - \Tpre$ observation times are referred to as the \emph{post-treatment} period.
We presume that the pre-treatment period is of size at least $\Tpre \geq 2$, so that at least one pre-treatment violation of parallel trends is well-defined.

We posit the existence of a population distribution $\popdist$ over a binary treatment indicator variable $D$ and real-valued potential outcomes $\poout{}{t}{1}$ and $\poout{}{t}{0}$ for time-periods $t =1 \dots T$.
Under the classical assumptions of consistency, no anticipation, and no interference, the observed outcome $\obsout{}{t}$ at time $t$ is related to the potential outcomes as $\obsout{}{t} = \poout{}{t}{1} \indicator{D=1 \textrm{ and } t \geq t_0} + \poout{}{t}{0} \indicator{D=0 \textrm{ or } t < t_0}$.
Our results apply both to a panel data setting, where $n$ subjects have been tracked over the $T$ time periods, or to repeated cross-sections, where for each of the $T$ time periods, we observe independent cross-sections comprising $n_t$ units.

The estimands considered in this paper are variants of the Average Treatment Effect on the Treated (ATT), which are commonly considered in the DID setting \citep{abadie_semiparametric_2005}.
We define the ATT in the $t$-th time-period as
\[
        \attt = \E{ \poout{}{t}{1} - \poout{}{t}{0} \lvert D = 1 } \enspace.
\]
While there are many estimands that may be of substantive interest to researchers, the majority of this paper focuses on the average post-treatment ATT, that is,
\begin{align}\nonumber
    \totalatt = \frac{1}{\Tpost} \sum_{t=t_0}^{T} \attt \enspace.
\end{align}
Our results can easily be extended to any linear function of the individual time-period ATTs $\attt$.
For example, this includes estimands based on averages which discount time-periods based on the time passed since treatment was administered.
See Appendix~\suppref{App:OtherEstimands} for more details.

Without further assumptions the ATT is not identified. 
However, the related difference-in-differences estimand
\[
    \ddt = \E{ \obsout{}{t} - \obsout{}{t_0-1} \mid D = 1} - \E{ \obsout{}{t} - \obsout{}{t_0-1}\mid D = 0 }
\]
is identified.
We define the corresponding average post-treatment difference-in-differences estimand as $\totaldd = \frac{1}{\Tpost} \sum_{t=t_0}^{T} \ddt$, which is also identified.
These two estimands $\totalatt$ and $\totaldd$ coincide under parallel trends and otherwise disagree based on the amount to which parallel trends are violated.
We have $\attt \equiv \ddt - \overviot$, 
where $\overviot$ is the \emph{overall violation} of the parallel trends at time $t$, i.e.
\[
    \overviot = \E{ \poout{}{t}{0} - \poout{}{t_0-1}{0} \mid D  = 1} - \E{ \poout{}{t}{0} - \poout{}{t_0-1}{0} \mid D = 0} \enspace.
\]
The overall violations $\overviot$ in the pre-treatment period $t \leq t_0-1$ are often estimated to test for violations of parallel trends (e.g., in the form of lead checks \citep{angrist_mostly_2009,egami_using_2023} or for visual inspection in an event-study plot \citep{gallagher_learning_2014,fitzpatrick_early_2014,markevich_economic_2018}).
For each time period $t$, we additionally define the \emph{iterative violation} of parallel trends $\iterviot$ as
\[
    \iterviot = \E{ \poout{}{t}{0} - \poout{}{t-1}{0} \mid D  = 1} - \E{ \poout{}{t}{0} - \poout{}{t-1}{0}  \mid D = 0 } \enspace.
\]
The iterative violations $\iterviot$ correspond to differences between two successive time periods, whereas the overall violations $\overviot$ correspond to differences in a given time period relative to a fixed reference $t_0-1$ time period.
The difference between overall and iterative violations appears only when the number of pre-treatment periods is $\Tpre > 2$ or the number of post-treatment periods is $\Tpost > 1$.
We find that reasoning about the iterative violations is often more intuitive because differences between successive time periods may often have a more substantive interpretation than differences to a fixed reference period.
In any event, the overall violations may be obtained from the iterative violations, and vice versa, according to the relations
\begin{align}\nonumber
\overviot &= \sum_{s=t_0}^t \itervio{s} \hspace{1.2cm} (t \geq t_0), \\ \nonumber
\overviot &= -\sum_{s=t+1}^{t_0-1} \itervio{s} \hspace{0.5cm} (t \leq  t_0 -1) 
\quad \text{and} \quad \\\nonumber
 \iterviot &= \overvio{t} - \overvio{t-1} \enspace.
\end{align}

In this way, our theoretical results can be phrased in terms of either notion of violation, i.e., see Appendix~\suppref{App:AbsoluteViolations}. 
We discuss the choice of notion of violation and the consequences for the results of our method in Section \ref{Sec:GunLaw} using the effects of right-to-carry laws as an illustration.

\subsection{Estimators}

Our results apply to a broad class of estimators with mild assumptions on the data generating process.
For example, our results will allow both for estimators which are simple comparisons of sample means as well as estimators which use covariate adjustment for improved precision.
We presume that the analyst has used the observed data to construct estimators of the two types of identifiable population quantities described above.
In particular, we assume that for each time in the pre-treatment period $t \in 2 \dots t_0 - 1$, the analyst has constructed an estimator $\estiterviot$ for the iterative parallel trend violations $\iterviot$.
Likewise, we assume that for each time in the post-treatment period $t \in t_0 \dots T$, the analyst has an estimator $\estddt$ for the difference-in-differences estimand $\ddt$.
The natural estimator for the average difference-in-differences across the post-treatment period is given by $\esttotaldd = \frac{1}{\Tpost} \sum_{t=t_0}^T \estddt$.
We group the population quantities and the corresponding estimators into a $T-1$ length vectors $\popfunc$ and $\estpopfunc$ defined as
\[
\popfunc = \paren[\Big]{ \itervio{2} , \dots \itervio{t_0-1}, \dd{t_0}, \dots \dd{T} }
\quadand
\estpopfunc = \paren[\Big]{ \estitervio{2},  \dots \estitervio{t_0-1}, \estdd{t_0}, \dots \estdd{T} }
\enspace,
\]
respectively.
We emphasize that this vector $\popfunc$ is a functional of the population distribution $\popdist$ rather than a parameter; in fact, no parametric assumptions on $\popdist$ will be made.
We also assume that the analyst has constructed an estimator $\estcov$ for the covariance of the estimator $\estpopfunc$. 

\subsection{Asymptotic Analysis}

We analyze the proposed statistical methods using an asymptotic framework.
The primary benefit of employing asymptotic analysis is that it will allow the functional $\popfunc$ to depend on the sample size $n$.
In this way, we will be able to formally speak about parallel trend violations which are ``present but negligible'' (e.g. on the order of $n^{-1/2}$) in contrast to those which are ``non-negligible'' (e.g. on the order of $n^{-1/4}$).
Formally, we will consider a sequence of population distributions indexed by the sample size, $\setb{\popdist_n}_{n=1}^\infty$.
Consequently, this yields a sequence of functionals $\setb{\popfunc_n}_{n=1}^\infty$ and estimators $\setb{ \estpopfunc_n }_{n=1}^\infty$ which are also indexed by the sample size.
In our asymptotic analysis, we treat the size of the time periods $T$ as being fixed as the sample size $n$ grows.

We place two assumptions on the data generating process. 
The first asymptotic assumption is that the errors of the estimators are jointly normal in large samples.
The second assumption is that the covariance estimator is consistent.

\begin{assumption}[Asymptotic Normality] \label{a:uniform_asymptotic_normality}
	The sequence of functionals converges and the estimators are asymptotically jointly normal: $\popfuncn \to \popfunc$ and $\sqrt{n} \cdot \paren[\big]{ \popfuncn - \estpopfuncn }
	\xrightarrow{d}
	\mathcal{N}(0,\pfcov)$.
\end{assumption}

\begin{assumption}[Consistent Variance Estimation] \label{a:unifConsistency}
    The variance estimator satisfies $\pfcov -\estcov_n  \xrightarrow{p} 0$ and $\pfcov$ is non-singular.
\end{assumption}

These two assumptions on the estimators and data generating process are quite mild.
For example, when the analyst uses simple sample mean estimators for the population quantities, then Assumptions~\ref{a:uniform_asymptotic_normality} and \ref{a:unifConsistency} follow from existence of the fourth moments of potential outcomes in the population distribution $\popdist$.

For positive sequences $x_n$ and $y_n$, we write $x_n = \bigO{y_n}$ if $\limsup x_n / y_n < \infty$, $x_n = \bigOmega{y_n}$ if $\liminf x_n / y_n > 0$, and $x_n = \bigTheta{y_n}$ if $x_n = \bigO{y_n}$ and $x_n = \bigOmega{y_n}$.
Similarly, we write $x_n = \littleO{y_n}$ if $\liminf x_n / y_n = 0$ and $x_n = \littleOmega{y_n}$ if $\limsup x_n / y_n = \infty$.
We remark that the proofs of all formal statements in this paper appear in the supplementary materials.
	
	\section{The Conditional Extrapolation Assumption}
\label{Sec:ExtrapolationAssumption}

In this section, we provide a formal description of the conditional extrapolation assumption stated informally in the introduction.
We study its implication for the bias of the difference-in-differences estimator.

\subsection{Extrapolating the Violations of Parallel Trends}

We begin by introducing a measure of the extent to which the parallel trends assumption is violated.
For a fixed $p \geq 1$, the \emph{severity of the parallel trends violations} in the pre-treatment and post-treatment periods,  denoted $\Spre$ and $\Spost$ are defined as
\[
\Spre = \paren[\Bigg]{ \frac{1}{\Tpre-1} \sum_{t = 2}^{t_0-1} \abs[\big]{ \iterviot }^p }^{1/p}
\quadand
\Spost= \paren[\Bigg]{ \frac{1}{\Tpost} \sum_{t = t_0}^{T} \abs[\big]{ \iterviot }^p }^{1/p}
\enspace,
\]
respectively.
These quantities measure the magnitude of the iterative violations in each of the periods.
Both $\Spre$ and $\Spost$ are increasing for larger values of $p$.
For example, $p=1$ is a simple average of the absolute iterative violations within period, while $p=\infty$ yields the maximum absolute iterative violation within a period.

The iterative violations $\iterviot$ in the pre- and post-treatment periods are population quantities and so too are the severity measures $\Spre$ and $\Spost$.
The analyst can estimate the violations in the pre-treatment period and so $\Spre$ can be consistently estimated using the observed data.
On the other hand, $\Spost$ is unidentified and therefore cannot be estimated from observed data.
The conditional extrapolation assumption posits a relationship between $\Spre$ and $\Spost$, when $\Spre$ satisfies a certain condition:

\begin{assumption}[Conditional Extrapolation] \label{a:extrapolation}
	If $\Spre \leq M$, then $\Spost \leq \Spre$.
\end{assumption}

The conditional extrapolation assumption states that if the violation of parallel trends is below an acceptable threshold $\Spre \leq M$, then extrapolation of these violations to the post-treatment period is justified, i.e. $\Spost \leq \Spre$.
On the other hand, if the violations of parallel trends in the pre-treatment period are larger than $M$, then no extrapolation is possible, i.e. no upper bound on $\Spost$ can be justified.
We refer to the condition that $\Spre \leq M$ as the \emph{extrapolation condition}.

Perhaps the strongest justification for the conditional extrapolation assumption is that it gives credence to the wide-spread use of preliminary testing in DID analyses.
Under the conditional extrapolation assumption, the first step of a DID analysis will be to verify or refute the extrapolation condition, i.e. that $\Spre \leq M$.
In contrast, the parallel trends assumption says nothing directly of the trends in the pre-treatment period and so testing them is irrelevant and even harmful for inference of the causal effect \citep{roth_pretest_2022}.
In a similar manner, more conventional partial identification assumptions, where violations of parallel trends in the pre-treatment period are assumed to unconditionally extrapolate to the post-treatment period, do not necessitate any type of preliminary testing \citep{manski_how_2018,rambachan_more_2023}.
In other words, the conditional extrapolation assumption is justifiable because it validates current best practices in applied research.

The conditional extrapolation assumption may also be justified on epistemological grounds.
Before carrying out a DID analysis, the applied researcher typically has (either explicitly or implicitly) a substantively plausible mechanism which implies that treatment and control groups would evolve similarly in the absence of treatment.
For example, the plausible mechanism in the border continuity design of \citet{Card_Krueger_94} is that neighboring counties evolve via the same economic dynamics in response to the same economic shocks.
If the mechanism is considered to be correct, then its implication holds and the DID analysis is warranted.
The trouble is that the mechanism, although plausible, cannot itself be empirically verified.
One epistemological approach is to attempt to falsify the plausible mechanism by investigating the extent to which its implications hold in the pre-treatment period.
If pre-treatment trends are sufficiently dissimilar, then the explanation of the mechanism is rejected and the DID analysis is deemed inappropriate.
On the other hand, if the pre-treatment trends are acceptably similar, then the explanation of the mechanism seems plausible and the DID analysis is carried out.
This line of reasoning is the epistemological essence which we have attempted to formalize in the conditional extrapolation assumption.

While we see the conditional extrapolation assumption as being justifiable in many cases, we do not claim that it is more or less verifiable than the parallel trends assumption.
Indeed, the conditional extrapolation assumption is counterfactual in nature: it places assumptions on the unidentifiable means $\E{ \poout{}{t}{0} \mid D=1 }$.
In this way, the validity of both the parallel trends assumption and the conditional extrapolation assumption lie beyond the purview of empirical scrutiny. 
For this reason, neither assumption can be considered more or less ``correct'', at least from the perspective of being empirically verifiable. 
Many of the concerns with the classical parallel trends assumption, such as endogenous treatment timing and anticipatory effects, still apply to the conditional extrapolation assumption. 
Running the preliminary test we describe in this paper does not absolve the analyst from reasoning about these types of issues.

A central question for applied researchers using Assumption~\ref{a:extrapolation} will be how to choose $p$, which determines how violation severity is measured, and $M$, which determines the threshold for violation severity under which extrapolation from the pre- to post-treatment period is warranted.
It is important to highlight that both the choice of $M$ and $p$ crucially rely on the application at hand, subject knowledge about the phenomenon under study, and the analyst's beliefs about when extrapolation is justified.
For this reason, we cannot give a one-size-fits-all recommendation, but we can highlight a few aspects of what such a decision entails.

The choice of $p$ reflects how the violations are aggregated across the times throughout the period.
The analyst may believe that each individual violation in the post-treatment period could be as extreme as the largest individual violation in the pre-treatment period, in which case $p = \infty$ is an appropriate choice.
In some situations, bounding each post-treatment violation of parallel trends by the most extreme violation in the pre-treatment period may be too stringent.
If, on the other hand, a singular severe deviation in the pre-treatment period is judged to be an outlier or extremely rare, $p=1$ may be the preferred choice as it measures severity by averaging all violations within the period.
Intermediate values of $p$ between $1$ and $\infty$ trade off between these two perspectives.

The choice of $M$ reflects the maximum severity under which extrapolation is justified.
Depending on the particular phenomenon under study, researchers need to gauge what amount of violation of parallel trends in the pre-treatment period (if any) would make extrapolation to the post-treatment be plausible. 
Researchers should choose $M$ so that extrapolation for violations $\Spre \leq M$ is justifiable whereas extrapolations for violations $\Spre \geq M$ would appear dubious.

\subsection{Implications for Inference}

Analysts are only able to estimate what is identifiable; in this case, that will be  $\totaldd$, the average of the difference-in-differences estimand in the post-treatment period.
If the parallel trends assumption were to hold exactly, then $\totaldd$ will correspond exactly to the causal estimand $\totalatt$, the average ATT in the post-treatment period.
Our goal in this section is to understand how far apart these two quantities{\textemdash}one identified, one not{\textemdash}will be under the conditional extrapolation assumption (Assumption~\mainref{a:extrapolation}).

\expcommand{\biasbound}{
	Suppose conditional extrapolation (Assumption~\mainref{a:extrapolation}) holds.
	If $\Spre \leq M$, then the difference between the average post-treatment DID and average post-treatment ATT is bounded as
	\[
	\abs[\big]{ \totalatt - \totaldd }
	\leq \kappa \cdot \Spre \enspace,
	\]
	where $\kappa = \paren{\frac{1}{T_{post}} \sum_{t=1}^{T_{post}}t^q}^{1/q}$ for $q$ satisfying $\frac{1}{q}+\frac{1}{p} =1$.
	If $\Spre > M$, then $\abs{ \totalatt - \totaldd }$ may be arbitrarily large.
}
\begin{proposition} \label{prop:bias}
	\biasbound
\end{proposition}

Proposition~\ref{prop:bias} highlights the role of the acceptable threshold $M$ in the extrapolation.
If the extrapolation condition is true, then we can produce a sharp upper bound on the difference between the causal estimand $\totalatt$ and the identified estimand $\totaldd$; otherwise, extrapolation is not justified and the two quantities may be arbitrarily far apart.
Already we can see that it will be of vital importance to test the validity of the extrapolation condition (i.e. $\Spre \leq M$) when conducting any statistical inference on the causal estimand $\totalatt$.
In Supplement~\suppref{supp:technical-lemmas}, we show that the upper bound in Proposition~\ref{prop:bias} is sharp.

When the extrapolation condition is true, the upper bound in Proposition~\ref{prop:bias} is the product of two terms, $\Spre$ and $\extrapbias$.
The first term $\Spre$ is the severity of parallel trend violations in the pre-treatment period.
If the severity $\Spre$ in the pre-treatment period is larger, then the possible difference between the causal and identified estimands becomes larger.
This is to be expected, given that the conditional extrapolation assumption posits that $\Spre$ is a bound on the severity $\Spost$ in the post-treatment period, when the extrapolation condition is true.
The second term in the upper bound is $\extrapbias$, which captures dependence on the length of the post-treatment period $\Tpost$ and the value $p$ used to define the severity.
Note that $\extrapbias$ is increasing with $\Tpost$, which reflects the fact that extrapolation becomes more challenging when the length of the post-treatment period increases, and decreasing with $p$.
In our asymptotic analysis, both $T$ and $p$ remain fixed, so that $\extrapbias$ is treated as being constant as the sample size $n$ grows.

Because the quantity $\extrapbias$ is known and $\Spre$ is identifiable, it is possible to estimate the sharp bound on the bias in Proposition~\ref{prop:bias}.
We will return to this idea in later sections.
The following corollary uses Proposition~\ref{prop:bias} to illustrate the error of the aggregated difference-in-differences estimator.

\begin{corollary} \label{corollary:interpret-bias}
	Suppose the CLT and conditional extrapolation (Assumptions~\ref{a:uniform_asymptotic_normality} and \ref{a:extrapolation}) hold.
	If $\Spre \leq M$, then the error of the DID estimator for the average post-treatment ATT is bounded as
	\[
	\totalatt - \esttotaldd
	= \bigOp[\Big]{ \frac{1}{\sqrt{n}} + \kappa \cdot \Spre }
	\enspace.
	\]
	Otherwise, if $\Spre > M$, then $\abs{\totalatt - \esttotaldd}$ may be arbitrarily large.
\end{corollary}

Corollary~\ref{corollary:interpret-bias} bounds the magnitude of the statistical error in using $\esttotaldd$ to estimate $\totalatt$.
The magnitude of this statistical error will have implications for the width of confidence intervals constructed from $\esttotaldd$ and an estimate of its variance.
Here, we look ahead somewhat by discussing the magnitude of the statistical error to gain intuition about the width of confidence intervals.
We defer the precise description and analysis of the proposed confidence intervals to Section~\ref{Sec:Inference}.

If the extrapolation condition holds, then the error in Corollary~\ref{corollary:interpret-bias} has two terms.
The $n^{-1/2}$ term reflects the statistical error inherent in the estimator $\esttotaldd$.
The $\kappa \cdot \Spre$ term reflects the largest possible bias incurred by estimating the identifiable $\totaldd$ and not the (unidentified) causal estimand $\totalatt$.
Conventional confidence intervals under the parallel trends assumption, where $\totalatt$ is identified, will have width of order $n^{-1/2}$.
Under our extrapolation assumption, the statistical error{\textemdash}and thus, the width of intervals{\textemdash}will be on the order $\max \setb{ n^{-1/2}, \Spre }$.
On the one hand, if the violations in the pre-treatment period are large, e.g. $\Spre = \bigOmega{1}$, then intervals will have constant width.
On the other hand, if the violations are ``present but negligible'', e.g. $\Spre = \bigO{n^{-1/2}}$, then the width of intervals will decrease at the same rate as conventional intervals.

	\section{Testing for Pre-Treatment Violations of Parallel Trends}
\label{Sec:Testing}

Assumption~\ref{a:extrapolation} states that if the pre-treatment violations of parallel trends are below an acceptable level, then the extrapolation of violations from the pre-treatment period to the post-treatment period is justified.
As shown in Section~\ref{Sec:ExtrapolationAssumption}, determining whether the extrapolation condition is true will be of crucial importance when invoking the conditional extrapolation assumption.
In this section, we provide a simple preliminary test for this purpose and evaluate its asymptotic properties.

\subsection{Asymptotic Consistency}

We begin by defining the notion of an asymptotically consistent preliminary test.
Informally speaking, a statistical test is asymptotically consistent if the Type I and Type II errors over a large class of null and alternative hypotheses go to zero as the sample size increases \citep{Robins_2003,shao_mathematical_2007,Lehmann_Romano_2022}.
Given a sequence $\setb{s_n}_{n=1}^\infty$, we define the \emph{$s_n$-separated null and alternative hypotheses} $H_0$ and $H_1$ as
\begin{align*}
	H_0(s_n) &= \setb[\big]{ \textrm{ Sequences } \popdist_n \textrm{ satisfying Assumption~\ref{a:uniform_asymptotic_normality} and } \Spre \leq M - s_n } \\
	H_1(s_n) &= \setb[\big]{  \textrm{ Sequences } \popdist_n \textrm{ satisfying Assumption~\ref{a:uniform_asymptotic_normality} and } \Spre \geq M + s_n }
	\enspace.
\end{align*}
$H_0$ contains data generating processes for which the pre-treatment violations $\Spre$ are less than $M - s_n$, so that the extrapolation condition is true.
Likewise, $H_1$ contains data generating processes for which the pre-treatment violations $\Spre$ are larger than $M + s_n$, in which case the extrapolation condition is false.
We allow for the pre-treatment violations $\Spre$ and the acceptable level $M$ to depend on the sample size $n$, but we suppress this notation for clarity.

The sequence $s_n$ represents the separation between the null and alternative hypotheses.
Some separation is required because no preliminary test will be able to have small Type I and Type II error when, for a given sample size, $\Spre$ is arbitrarily close to $M$.
It is, however, possible to construct tests with vanishing Type I and Type II error when the difference $\abs{\Spre - M}$ goes to zero as the sample size increases, albeit at a sufficiently slow rate.
In this case, the preliminary test can be interpreted as having vanishing Type I and Type II error for hypotheses where $\Spre$ is arbitrarily close to $M$, provided that the sample size is sufficiently large.

A preliminary test will be a function $\phi$ which maps the observed data to $\setb{0,1}$.
If $\phi$ evaluates to $1$ on the data, then the null hypothesis that $\Spre \leq M$ is rejected and the extrapolation condition is declared to be false.
If $\phi$ evaluates to $0$ on the data, then the test fails to reject the null hypothesis that $\Spre \leq M$, and the extrapolation condition is declared to be true.

A preliminary test will be \emph{asymptotically consistent for separation $\setb{s_n}_{n=1}^\infty$} if
\[
\lim_{n \to \infty} \Esub[\big]{X \sim \popdist_n^{(0)} }{ \phi(X)} = 0
\quadand
 \lim_{n \to \infty} \Esub[\big]{X \sim \popdist_n^{(1)} }{ 1 - \phi(X)} = 0
\]
for all sequences $\popdist_n^{(0)} \in H_0(s_n)$ and $\popdist_n^{(1)} \in H_1(s_n)$, where $X$ denotes the observed data.
Note that the first term is the limiting Type I error under the null and the second term is the limiting Type II error under the alternative.
For an asymptotically consistent preliminary test, both of these limiting errors will be zero for all data generating processes in the (separated) null and alternative.

\subsection{A Simple Preliminary Test}

We now provide a simple preliminary test which is shown to be asymptotically consistent.
Recall that the severity of parallel trend violations in the pre-treatment period is defined as $\Spre = \paren{ (\Tpre-1)^{-1} \sum_{t=0}^{t_0-1} \abs{\iterviot}^p }^{1/p}$, where $\iterviot$ are the iterative violations of parallel trends.
The analyst has already constructed estimators $\estiterviot$ for these iterative violations.
We define the \emph{estimated severity of pre-treatment violations} $\estSpre$ as
\[
\estSpre = \paren[\Big]{ \frac{1}{\Tpre-1} \sum_{t=2}^{t_0-1} \abs[\big]{ \estiterviot  }^p }^{1/p} \enspace,
\]
which substitutes the iterative violations $\iterviot$ with their estimates $\estiterviot$.
We propose the preliminary test $\phi$ which checks whether the estimated severity of pre-treatment violations is below the acceptable level, i.e.
\[
\phi = \indicator{ \estSpre \geq M }
\enspace.
\]
This test is intuitive and simple to implement.
The following theorem shows that it is asymptotically consistent under a separation sequence that goes to zero sufficiently slowly.

\expcommand{\testingthm}{%
	The preliminary test $\phi$ described above is asymptotically consistent for separation $s_n = \littleOmega{n^{-1/2}}$.
}
\begin{theorem} \label{thm:testing}
	\testingthm
\end{theorem}

The proof of Theorem~\ref{thm:testing} can be roughly explained as follows:
by Assumption~\ref{a:uniform_asymptotic_normality}, the individual estimators have errors $\iterviot - \estiterviot = \bigOp{n^{-1/2}}$ so that by an application of the continuous mapping theorem, $\Spre - \estSpre = \bigOp{n^{-1/2}}$.
If the separation is at least $s_n = \littleOmega{n^{-1/2}}$, then the estimator $\estSpre$ will be on the correct side of the acceptable threshold $M$ with probability tending to 1.

 Unlike most conventional statistical tests, our proposed preliminary test of the extrapolation condition is comparatively simple.
For example, our test did not require the derivation of a limiting distribution nor the specification of a critical value.
The reason is that the goal of our preliminary test{\textemdash}namely, asymptotic consistency{\textemdash}is different from the primary goal underlying most conventional statistical tests, which is to minimize Type II error subject to a fixed Type I error.
This latter goal is appropriate when the analyst expects to be in regimes where such a trade-off is required.
By imposing a separation between the null and the alternative on which asymptotic consistency is being defined, such a trade-off is no longer required and simpler tests are permitted.

We believe that constructing preliminary tests from the viewpoint of asymptotic consistency is appropriate when considering the conditional extrapolation assumption.
As a result, some type of separation condition is unavoidable.
As we shall see in Section~\ref{Sec:Inference}, this means that our inference will only be valid for data generating processes which satisfy the required separation condition.
We defer an extended discussion on the merits and drawbacks of this approach, as well as comparisons to existing tests for parallel trends, to Section~\ref{sec:comparison-to-alternative-intervals}.
	
	\section{Conditionally Valid Inference}
\label{Sec:Inference}

If the extrapolation condition is declared to be true, then extrapolation is justifiable under Assumption~\ref{a:extrapolation} and thus the analyst may proceed to perform statistical inference for the causal estimand.
In this section, we provide confidence intervals for the causal estimand $\totalatt$ which are \emph{conditionally valid}, which is to say that the intervals will cover at the nominal rate conditioned on the event that the preliminary test declares the extrapolation condition to be true.

\subsection{Confidence Intervals}

We propose confidence intervals which will asymptotically contain the causal estimand $\totalatt$ with probability at least $1 - \level$, when conditioned on the event that the preliminary test declares that the extrapolation condition is true.
Formally, we define the confidence interval $\ci$ as
\begin{equation} \label{Eq:CI}
       \ci = 
        \esttotaldd
        \pm 
        \braces[\Big]{
        	\extrapbias \cdot \estSpre
        	+ \frac{ f(\level, \estcov) }{ \sqrt{n}}
    	} \enspace.
\end{equation}

The intervals are simple to implement and do not require any significant computational resources beyond calculation of estimated quantities.
The confidence interval depends on three estimated quantities: the estimated average post-treatment difference-in-differences $\esttotaldd$, the estimated pre-treatment violations $\estSpre$, and the estimated covariance matrix $\estcov$.
Recall that the term $\extrapbias$ is completely known, depending only on the length of the post-treatment period $\Tpost$ and the quantity $p$ used to define the violation severity measures.
As we shall discuss later, the term $f(\alpha, \estcov)$ corresponds to a critical value of an appropriately chosen  limiting distribution and its evaluation is straightforward.

Let us now provide the intuition of the various parts of the confidence interval.
The intervals are centered at the point estimate $\esttotaldd$ of the average post-treatment difference-in-differences estimand, $\totaldd$.
In general, the difference-in-differences estimand $\totaldd$ is not equal to the causal estimand $\totalatt$, so $\esttotaldd$ will be a biased estimate for $\totalatt$.
For this reason, we inflate the interval by the factor $\extrapbias \cdot \estSpre$, which is an estimate of the sharp bound on this bias given by Proposition~\ref{prop:bias}.
Finally, we inflate the interval by an additional term $ f(\level, \estcov) \cdot n^{-1/2}$, which accounts for the statistical fluctuations in the joint estimates of $\esttotaldd$ and $\estSpre$.

More precisely, the function $f(\level, \covsig)$ describes the $(1-\alpha)$ critical values of the limiting distribution corresponding to statistical errors in $\esttotaldd$ and $\extrapbias \cdot \estSpre$.
To this end, consider the multivariate function $\psi : \Reals^{T-1} \to \Reals$ defined as
\[
\psi(x_1 \dots x_{T-1})  = \abs[\Big]{ \frac{1}{\Tpost} \sum_{t=T_{pre}}^{T-1} x_t }
+ \extrapbias \cdot \paren[\Big]{ \frac{1}{\Tpre-1} \sum_{t=1}^{T_{pre}-1} \abs{ x_t }^p }^{1/p}
\enspace,
\]
which is constructed to reflect the errors in $\esttotaldd$ and $\extrapbias \cdot \estSpre$.
For $t < t_0$, the variables $x_t$ correspond to errors in the estimated iterative violations $\iterviot - \estiterviot$ while for $t \geq t_0$, the variables $x_t$ correspond to errors in the estimated difference-in-differences $\ddt - \estddt$.
Assumption~\ref{a:uniform_asymptotic_normality} states that the errors of these estimators will be jointly well-approximated by a random vector $Z \sim \mathcal	{N}(0, \covsig)$.
We define the function $f(\level, \covsig)$ as satisfying the critical value, i.e. $\Pr{ \psi(Z) \geq f(\level, \covsig) } = \alpha$.
While the true covariance $\covsig$ is not known, Assumption~\ref{a:unifConsistency} implies that it suffices to substitute the estimated covariance $\estcov$ into this critical value.
Evaluating the critical value function $f(\level, \covsig)$ is simple, and we refer the reader to Section~\suppref{App:ProofPropCondInf} in the supplementary materials for details.

\expcommand{\conditionalcoverage}{%
	Suppose that Assumptions~\mainref{a:uniform_asymptotic_normality}-\mainref{a:extrapolation} hold.
	Then, the confidence interval $\ci$ is conditionally valid on the results of the preliminary test under the well-separated null; that is, if $P_n \in H_0(s_n)$ for $s_n = \littleOmega{n^{-1/2}}$, then
	\[
	\liminf_{n \to \infty } \Pr[\big]{ \totalatt \in \ci \mid \phi = 0 } \geq 1 - \level
	\enspace.
	\]
}

\begin{theorem} \label{theorem:cond-inf}
	\conditionalcoverage
\end{theorem} 

Theorem~\ref{theorem:cond-inf} guarantees that the proposed intervals are conditionally valid, in the sense that the coverage is at the nominal level when conditioned on the event that the preliminary test declares that the extrapolation condition is true.
The width of the confidence intervals will generally depend on the severity of parallel trend violations in the pre-treatment period.
The expected width of the intervals will scale as $\maxf{\Spre, n^{-1/2}}$.
If the violation of parallel trends in the pre-treatment period is large, then the expected width of the intervals will also be of the same order, e.g. if $\Spre = \bigOmega{1}$ then $\mathrm{width}(\ci) = \bigOmegap{1}$.
On the other hand, if the violation of parallel trends in the pre-treatment period is ``present but negligible'', then the width of the confidence intervals will decrease at the usual rate, e.g. if $\Spre = \bigO{n^{-1/2}}$, then $\mathrm{width}(\ci) = \bigOp{n^{-1/2}}$.

\subsection{Comparison to Alternative Confidence Intervals} \label{sec:comparison-to-alternative-intervals}

Several authors have shown that conventional confidence intervals are not conditionally valid on the results of conventional hypothesis tests for pre-treatment violations of parallel trends \citep[see e.g.,][]{freyaldenhoven_pre-event_2019,roth_pretest_2022}.
Practically speaking, this results in misleading conclusions about the causal estimand which can be severely biased in either direction. 

It may be instructive to illustrate why conventional confidence intervals suffer from these problems.
The main issue is that conventional hypothesis tests implicitly seek to detect any positive violations of the parallel trends assumption, i.e. acceptable level $M = 0$ and with separation $s_n = n^{-1/2}$.
As a result, conventional hypothesis tests must incur either a constant Type I or Type II error over this set of null and alternative hypotheses. 
Our proposed confidence intervals are able to overcome this issue by focusing only on achieving coverage in the well-separated null hypotheses, i.e. when $\abs{ \Spre - M } \geq \littleOmega{n^{-1/2}}$.

The confidence intervals proposed in this paper have both benefits and drawbacks.
The major benefit is that intervals are valid even after conditioning on the outcome of the preliminary test for violations of parallel trends.
The biggest drawback is that this validity of testing and inference holds only when the actual violation of parallel trends in the pre-treatment period $\Spre$ is sufficiently far from the acceptable threshold $M$.
Such a trade-off is necessary because testing is possible only up to some statistical error, which depends on the sample size.
We find this trade-off unproblematic in practical scenarios because we anticipate that the acceptable threshold $M$ will typically be moderately large and the difference $\abs{\Spre - M}$ will be larger than $n^{-1/2}$.

In Section~\ref{app:sens} of the appendix, we extend the proposed confidence intervals to accommodate sensitivity analysis to the form of the extrapolation.
In particular, we posit a more general conditional extrapolation assumption, parametrized by $\gamma \geq 1$, which stipulates that $\Spre \leq M$ implies that $\Spost \leq \gamma \cdot \Spre$.
We show how to conduct sensitivity analyses with respect to the choice of $\gamma$, by reporting the largest value $\gamma_\level^{\text{over}}$ for which the corresponding confidence interval does not contain 0.

\subsection{Probability of Valid Reporting} \label{sec:unconditional-coverage}

We have shown that the confidence intervals have asymptotically valid conditional coverage.
The property of conditional coverage has taken up much of the debate around the appropriateness of preliminary testing in the difference-in-differences literature.
However, researchers may also be interested in the probability of valid reporting.
In this section, we show that the probability of valid reporting for our proposed preliminary test and confidence interval achieves the nominal level asymptotically, under the same assumptions used to analyze conditional coverage.

The \emph{probability of valid reporting} is defined as
\[
\Pr{ \totalatt \in \ci \text{ and } \phi = 0 } \enspace,
\]
which is the probability that an interval is reported \emph{and} that it contains the truth.
While conditional coverage and the probability of valid reporting measure different quantities, they may both be relevant from a scientific perspective.
The conditional coverage may be more relevant from the critic's perspective (i.e. given that the preliminary test passed, is the confidence interval correct?) while the probability of valid reporting may be more relevant from the researcher or funder's perspective (i.e. is it likely that our study design will produce correct and reportable results?).
The following theorem shows when the extrapolation condition is true and the separation is sufficiently large, that the proposed confidence interval and test asymptotically attain a probability of valid reporting of at least the nominal level.

\expcommand{\unconditionalcoverage}{%
	Suppose that Assumptions~\mainref{a:uniform_asymptotic_normality}-\mainref{a:extrapolation} hold.
	Then, the probability of valid reporting for the confidence interval $\ci$ and test $\phi$ is of at least the nominal level under the well-separated null; that is, if $P_n \in H_0(s)$ for $s_n = \littleOmega{n^{-1/2}}$, then
	\[
	\liminf_{n \to \infty } \Pr[\big]{ \totalatt \in \ci \text{ and } \phi = 0 } \geq 1 - \level
	\enspace.
	\]
}

\begin{theorem} \label{theorem:uncond-inf}
	\unconditionalcoverage
\end{theorem}
	
	\section{Empirical Illustrations}

\subsection{Synthetic Numerical Simulations}

In this section, we conduct synthetic numerical simulations to empirically investigate characteristics of our proposed inference under the conditional extrapolation assumption. 
Specifically, we aim to answer two questions:
\begin{quote}
	\centering
	\textit{
	How does the width of our confidence interval $\ci$ scale with $\Tpre$ and $p$?
	\\
	What is the non-asymptotic conditional coverage and the probability of valid reporting?
	}
\end{quote}

While our theoretical results are asymptotic, the simulations provide an opportunity to investigate finite sample behavior.
In order to properly evaluate the proposed methodology, we focus on the most challenging regime when $\Spre$ is smaller than but close to the threshold $M$.

In our simulation, we construct a data generating process $\popdist$ which requires the specification of the following inputs: a level of violations of parallel trends $\Spre$ with parameter $p$, a threshold for acceptable levels of violations $M$, and the length of the pre- and post-treatment periods $\Tpre$ and $\Tpost$. 
According to these, we construct a population distribution $\popdist$ that satisfies the extrapolation assumption, from which we draw samples.

The first step in constructing the population $\popdist$ is to choose a post-treatment severity $\Spost$.
We choose $\Spost$ in a manner which is consistent with the conditional extrapolation assumption.
If $\Spre \leq M$, then we set $\Spost = \Spre$, otherwise we set $\Spost = 10 \cdot \Spre$.

The second step in constructing the population $\popdist$ is to specify the mean of all potential outcomes.
The expectations of the potential outcomes are chosen as follows:
\begin{align*}
	\E{ \poout{}{t}{0} \lvert D = 0 } &= \rho \cdot \E{ \poout{}{t-1}{0} \lvert D = 0 }
	+ \alpha \cdot  t + \log(T) \paren[\big]{ \cos(t) + \sin(t/2) } \\
	 \E{ \poout{}{t}{0} \lvert D = 1 } &= \E{ \poout{}{t}{0} \lvert D = 0 }  + \sum_{s=2}^t \itervio{s}  \\
	 \E{ \poout{}{t}{1} \lvert D = 1 } &= \E{ \poout{}{t}{0} \lvert D = 1 } + \beta \cdot \indicator{ D = 1 \text{ and } t \geq t_0 }.
\end{align*}
In the control group, the potential outcomes $\poout{}{t}{0}$ are chosen to exhibit a non-linear trend over time.
In the treatment group, the (unobserved) potential outcomes under control are chosen to have pre-specified iterative violations described below.
Finally, the treatment group is chosen to have treated potential outcomes which reflect a constant treatment effect (level-shift) for all post-treatment time-periods.
In all our simulations, we set $\alpha = 0.3$, $\rho = 0.7$ and $\beta = 2$.

We construct the iterative violations of parallel trends $\itervio{t}$ for time $t= 2 \dots T$ in the following manner.
First we construct preliminary values of the iterative violations as $\titervio{t}= \log(T) (\sin(t) + \cos(t/2))$, which are chosen in a non-linear fashion.
The severity measures of the preliminary iterative violations $s_{\mathrm{pre}} = \paren[]{ \frac{1}{\Tpre-1} \sum_{t = 2}^{t_0-1} \abs[\big]{ \titervio{t} }^p }^{1/p}$ and $s_{\mathrm{post}} = \paren[]{ \frac{1}{\Tpost} \sum_{t = t_0}^{T} \abs[\big]{ \titervio{t} }^p }^{1/p}$ will not necessarily match the specified levels $\Spre$ and $\Spost$.
In order to ensure that the constructed iterative violations match the specified severity levels $\Spre$ and $\Spost$, we simply scale the preliminary values to obtain the iterative violations: $\itervio{t} = \Spre/s_{\mathrm{pre}} \cdot \titervio{t}$ for all pre-treatment periods $t = 2,\dots t_0-1$ and $\itervio{t} = \Spost / s_{\mathrm{post}} \cdot \titervio{t}$ for all post-treatment periods $t=t_0 \dots T$.

Our sampling from the population $\popdist$ is chosen to imitate repeated cross-sections so that observed data will be independent across treatment groups and time.
We sample $n$ units for each combination of treatment group and time.
For units in the treatment group ($D_{i}=1$), we draw observations according to
\[Y_{i}^{(t)} \overset{iid}{\sim} \indicator{t \geq t_0 } \cdot\mathcal{N}(\E{ \poout{}{t}{1} \lvert D = 1 },\sigma_1^2) + \indicator{ t < t_0 }\cdot\mathcal{N}(\E{ \poout{}{t}{0} \lvert D = 1 },\sigma_1^2)
\]  
and for the control group ($D_{i}=0$), we draw from
\[Y_{i}^{(t)} \overset{iid}{\sim}  \mathcal{N}(\E{ \poout{}{t}{0} \lvert D = 0 },\sigma_0^2).
\]
We set $\sigma_1 =2.1$ and $\sigma_0 = 1.5$ for all simulations.
Estimation proceeds via non-parametric sample mean estimators, that is, 
\begin{align*}
\hat \tau_{DD}^{(t)} &= \frac{1}{n} \paren[\Big]{
	\braces[\Big]{ \sum_{i:D_{i} = 1} Y_{i}^{(t)} -\sum_{i:D_{i} = 1}Y_{i}^{(t_0-1)} } 
	- \braces[\Big]{\sum_{i:D_{i} = 0} Y_{i}^{(t)} -\sum_{i:D_{i} = 0}Y_{i}^{(t_0-1)}}
	} 
\quadand \\
\estiterviot &= \frac{1}{n} \paren[\Big]{
	\braces[\Big]{ \sum_{i:D_{i} = 1} Y_{i}^{(t)} -\sum_{i:D_{i} = 1}Y_{i}^{(t-1)} } 
	- \braces[\Big]{\sum_{i:D_{i} = 0} Y_{i}^{(t)} -\sum_{i:D_{i} = 0}Y_{i}^{(t-1)}}
	}
\enspace.
\end{align*}
In each of the simulations, reported expectations or probabilities are calculated using 5,000 Monte Carlo runs.

\begin{figure}[t]
	\centering  
    \begin{subfigure}{0.4\textwidth}
        \centering
        \includegraphics[width=\linewidth]{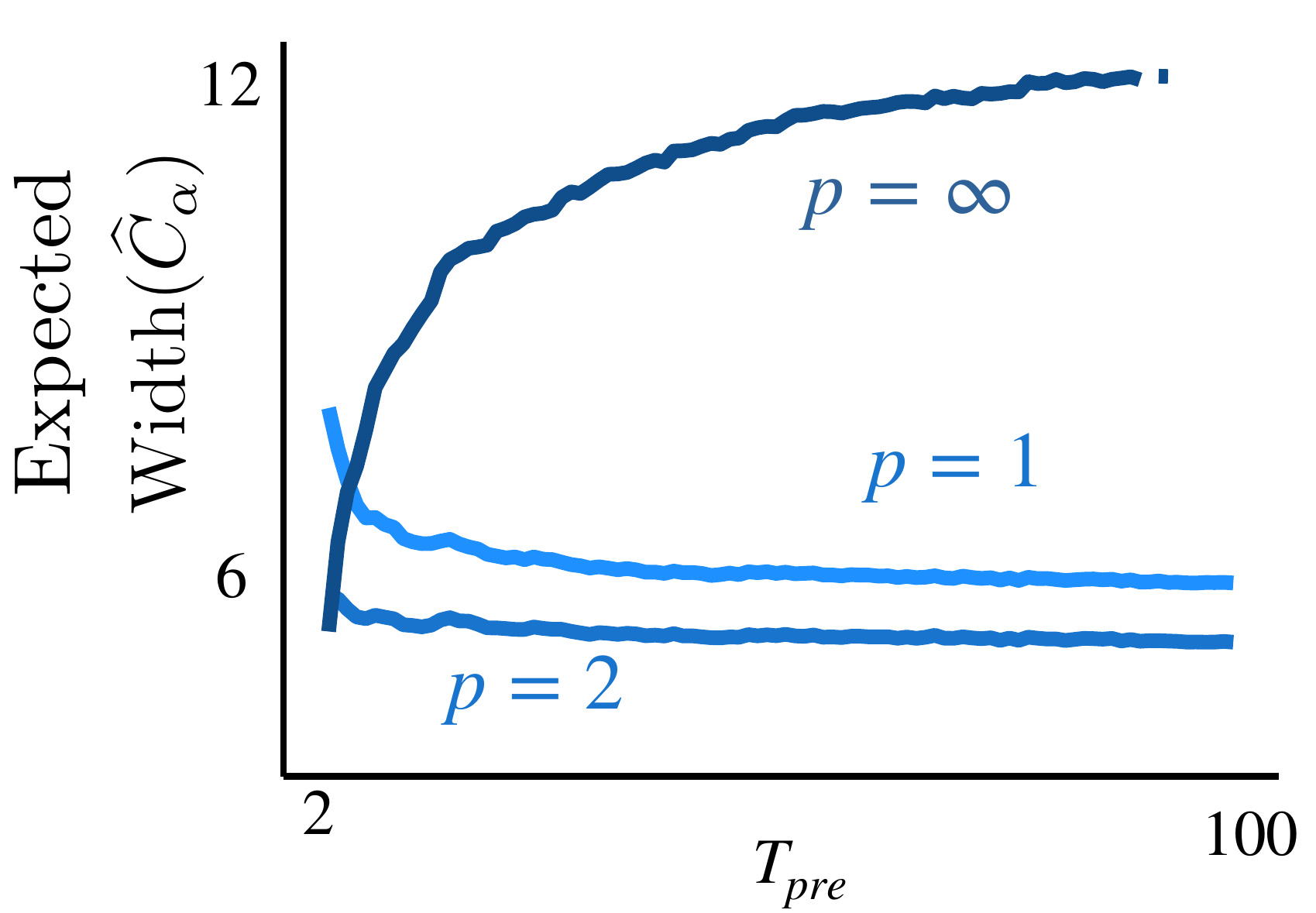}
        \caption{}
        \label{fig:Num_sim1_left}
    \end{subfigure}
    \hspace{1cm}
    \begin{subfigure}{0.4\textwidth}
        \centering
        \includegraphics[width=\linewidth]{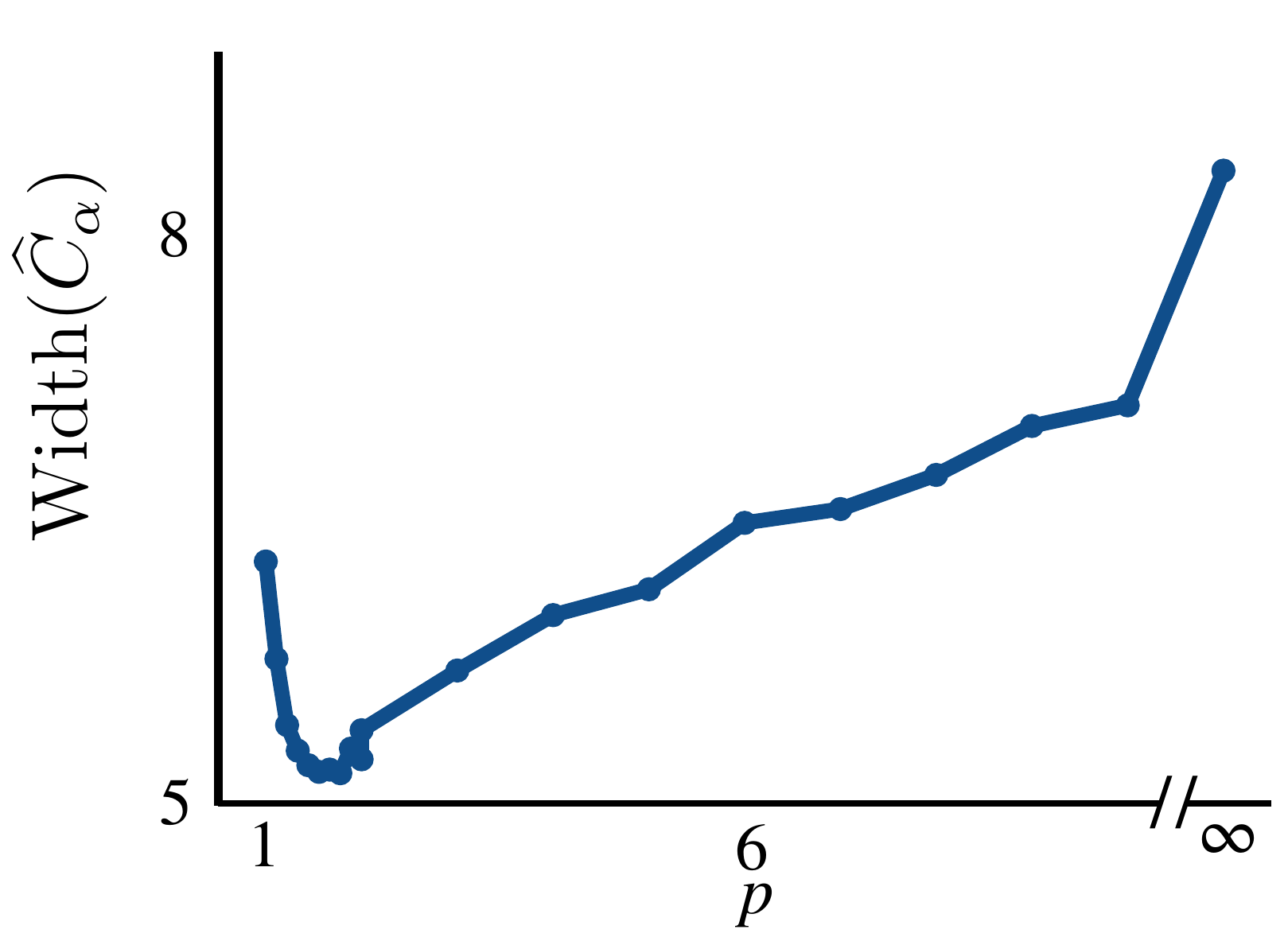}
        \caption{}
        \label{fig:Num_sim1_right}
    \end{subfigure}
	\caption{Characteristics of our confidence interval. (a) Expected width of our confidence interval as a function of the number of available pre-treatment periods $\Tpre$ for different values of $p$. (b) Confidence interval width as a function of $p$.
	}
	\label{fig:Num_sim1}
\end{figure}

\paragraph{Interval Width}
Our confidence intervals depend on both the available pre-treatment periods $\Tpre$ and the parameter of the severity measure $p$.
We consider a population where $\Spre = 0.1$, $M=100$, $\Tpost=4$, $n=100$, and we vary the number of pre-treatment periods $\Tpre$ from $2$ to $100$.
We begin by studying the impact of $\Tpre$ on the width of the intervals, using three values of the severity parameter $p \in \setb{1,2, \infty}$.
Figure \ref{fig:Num_sim1_left} shows the expected width of $\ci$ as a function of the number of pre-treatment periods $\Tpre$ included in the calculation of $\ci$.
For $p = \infty$, the width increases slowly with the number of pre-treatment periods $\Tpre$.
This growth is to be expected, given that $p=\infty$ measures the worst-case severity, which was chosen to slowly grow in our simulation.
For $p \in \setb{1,2}$, the width decreases with $\Tpre$, but eventually converges to some lower bound.
Next, we study the effect of $p$ on the interval by fixing $\Tpre=10$, drawing one sample, and varying $p$.
In Figure \ref{fig:Num_sim1_right}, we can see that the width of the confidence interval decreases for small values of $p$ before it begins to increase around $p \approx 2$.

\begin{figure}[t]
	\centering  
    \begin{subfigure}{0.30\textwidth}
        \centering
        \includegraphics[width=\linewidth]{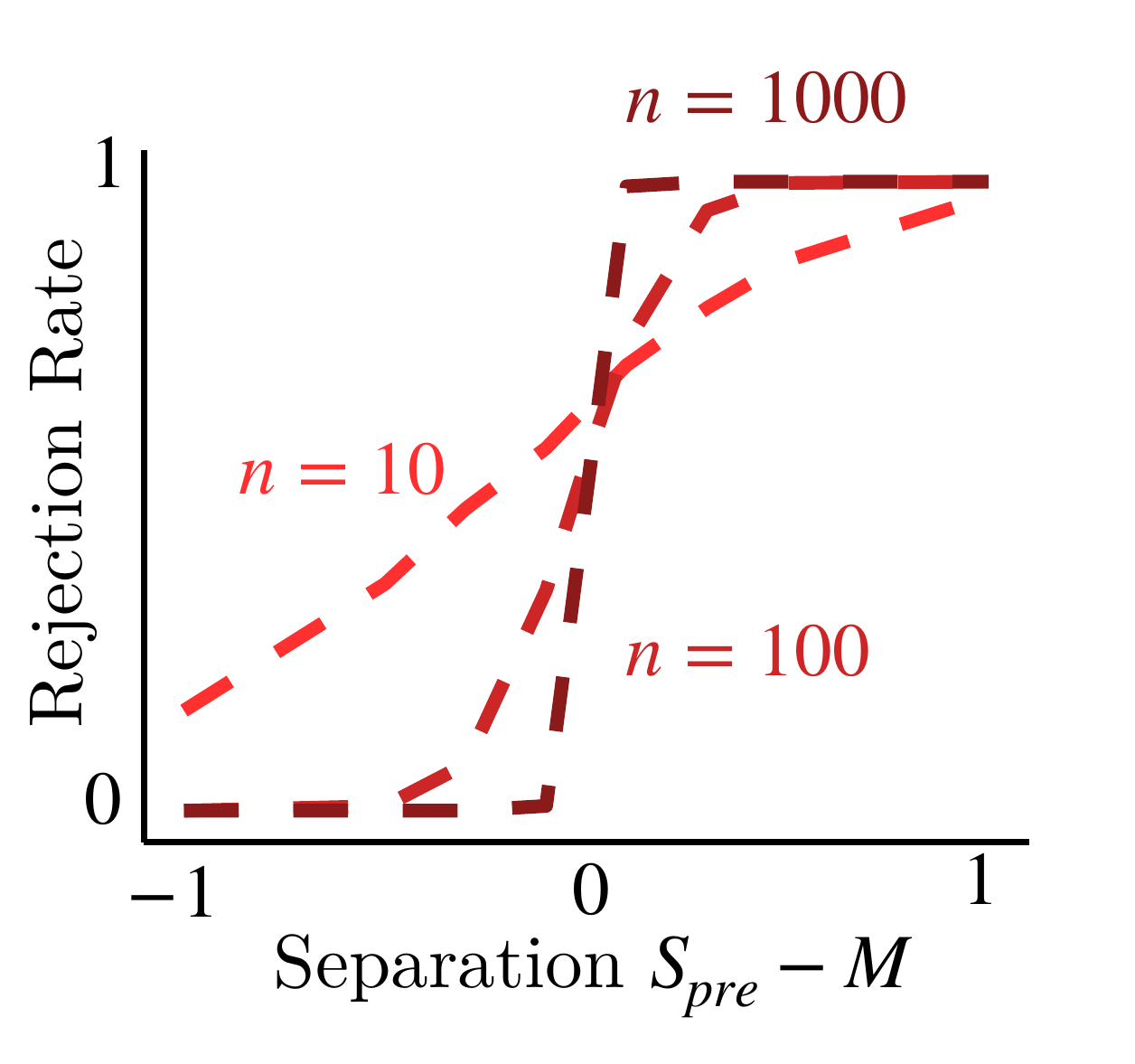}
        \caption{}
        \label{fig:Num_sim2_left}
    \end{subfigure}
    \hfill
    \begin{subfigure}{0.33\textwidth}
        \centering
        \includegraphics[width=\linewidth]{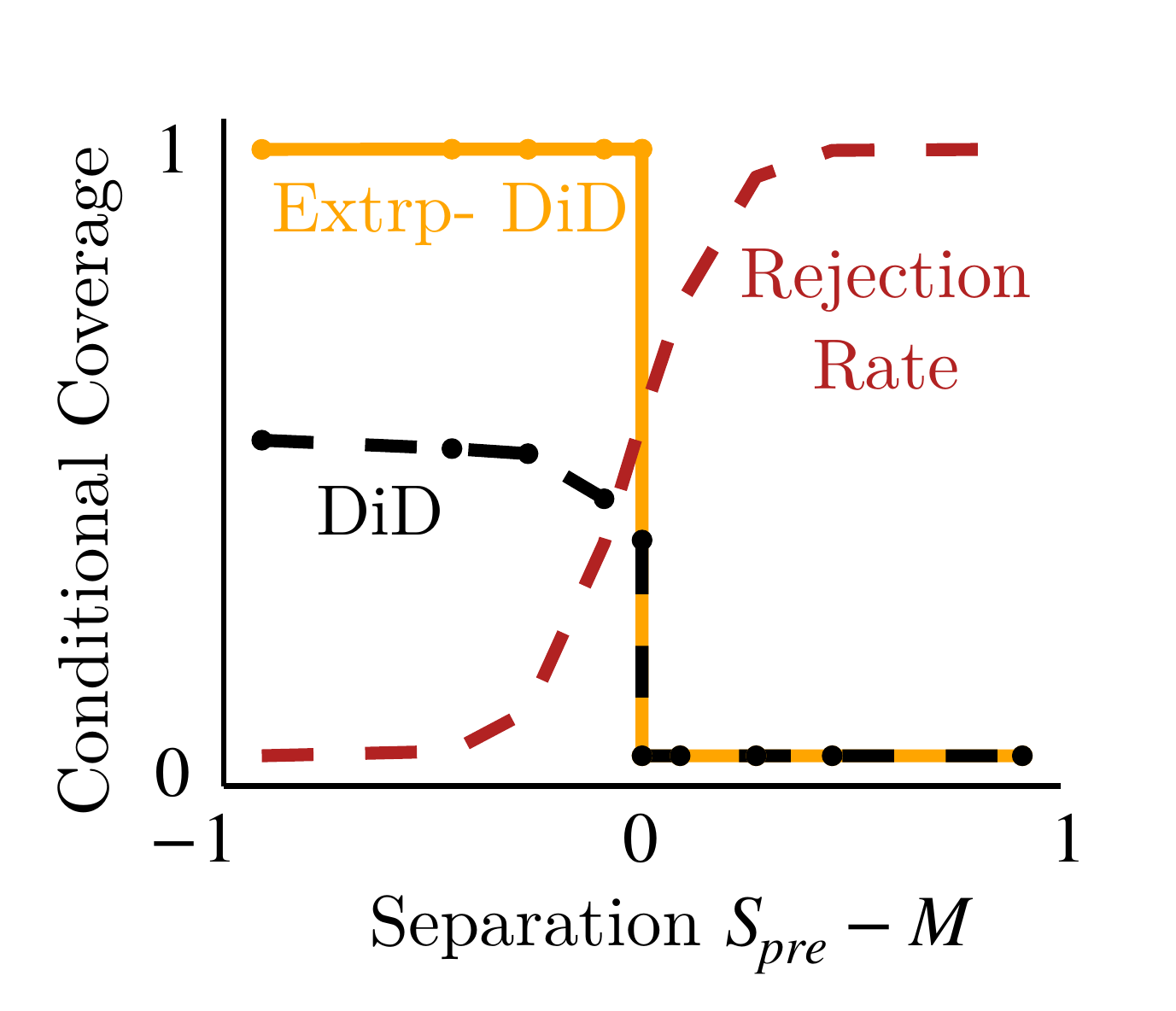}
        \caption{}
        \label{fig:Num_sim2_middle}
    \end{subfigure}
    \hfill
     \begin{subfigure}{0.33\textwidth}
        \centering
        \includegraphics[width=\linewidth]{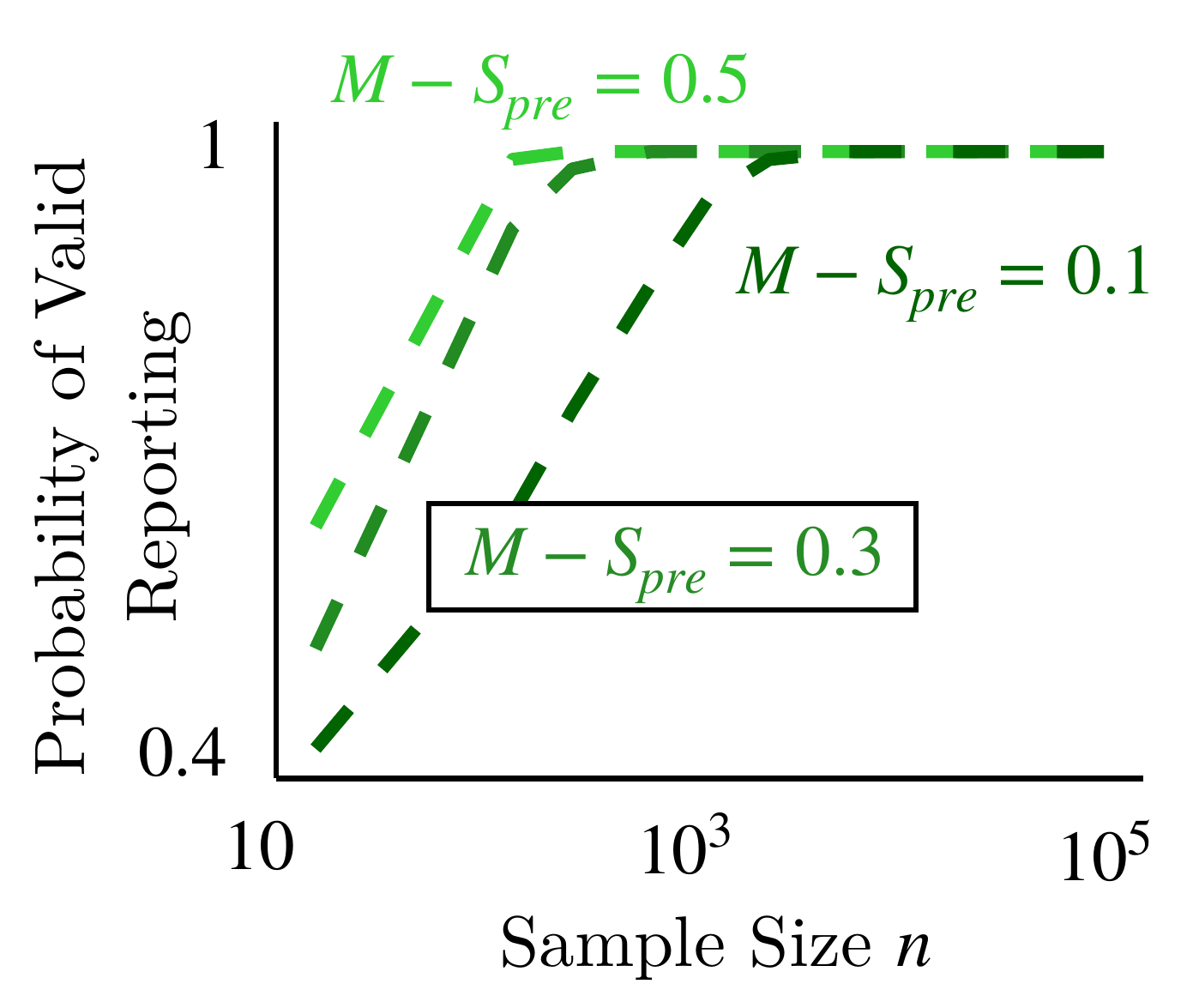}
        \caption{}
        \label{fig:Num_sim2_right}
    \end{subfigure}
	\caption{Effect of sample size on our test and conditional coverage.
		(a) Rejection rate of the preliminary test $\phi$ for three sample sizes $n$ as a function of the separation $\Spre -M$.  
		(b) Conditional coverage of conventional (black) and our (orange) confidence interval as a function of the separation $\Spre -M$. The dashed red line indicates the rejection rate of our test.
		(c) Probability of valid reporting for three levels of separation $M - \Spre$ as a function of sample size $n$.}
	\label{fig:Num_sim2}
\end{figure}

\paragraph{Interval Coverage}
In order to evaluate both the conditional coverage probability and the probability of valid reporting, we are interested in the challenging regime where the pre-treatment severity of parallel trend violations $\Spre$ is close to the acceptable threshold $M$.
Throughout this simulation, we fix $M = 2$, $\Tpre = 3$, and $\Tpost =1$, and we vary $\Spre$ to be closer or further from $M$.
We ensure that $\Spre \leq M$, so that the extrapolation is warranted.
We begin by comparing the rejection rates for different sample sizes $n$, see Figure \ref{fig:Num_sim2_left}. With increasing sample size, the rejection rate becomes steeper around the origin.
This indicates that the quality of the test improves as the sample size grows, as predicted by theoretical analysis.

Next, we evaluate conditional coverage by varying $\Spre$ so that $\Spre - M$ takes values in $[-1, 1]$.
The extrapolation condition is true when this value is negative and false when it is positive.
In Figure~\ref{fig:Num_sim2_middle}, we fix a sample size of $n=100$ and plot the conditional coverage of our proposed confidence interval (orange), the conditional coverage of the conventional interval (black), and the rejection rate (red).
We find that when the extrapolation condition is true, our intervals conditionally cover at the nominal level while the conventional intervals undercover.
When the extrapolation condition is false, none of the intervals have valid conditional coverage -- by construction of the population, $\Spost \gg \Spre$ when $\Spre > M$, so that the bias of the difference-in-differences estimator is severely underestimated.

We end by evaluating the probability of valid reporting as shown in Figure~\ref{fig:Num_sim2_right}, where we plot $\Pr{ \totalatt \in \ci \text{ and } \phi = 0 }$ for increasing sample size $n$ and have chosen three values of $\Spre$ so that $M - \Spre$ takes values $\setb{0.1, 0.3, 0.5}$.
We find that when $M$ and $\Spre$ are closer, the probability of passing the pretest and reporting a valid confidence interval is smaller.
Moreover, we find that for every fixed separation $M - \Spre$, the probability of valid reporting reaches the nominal level for a sufficiently large sample size, as predicted by theoretical analysis.

\subsection{Semi-Synthetic Illustration:  Effect of Right-to-Carry Laws on Crime Rates}
\label{Sec:GunLaw}
\begin{figure}[t] 
        \centering
        \includegraphics[width=\linewidth]{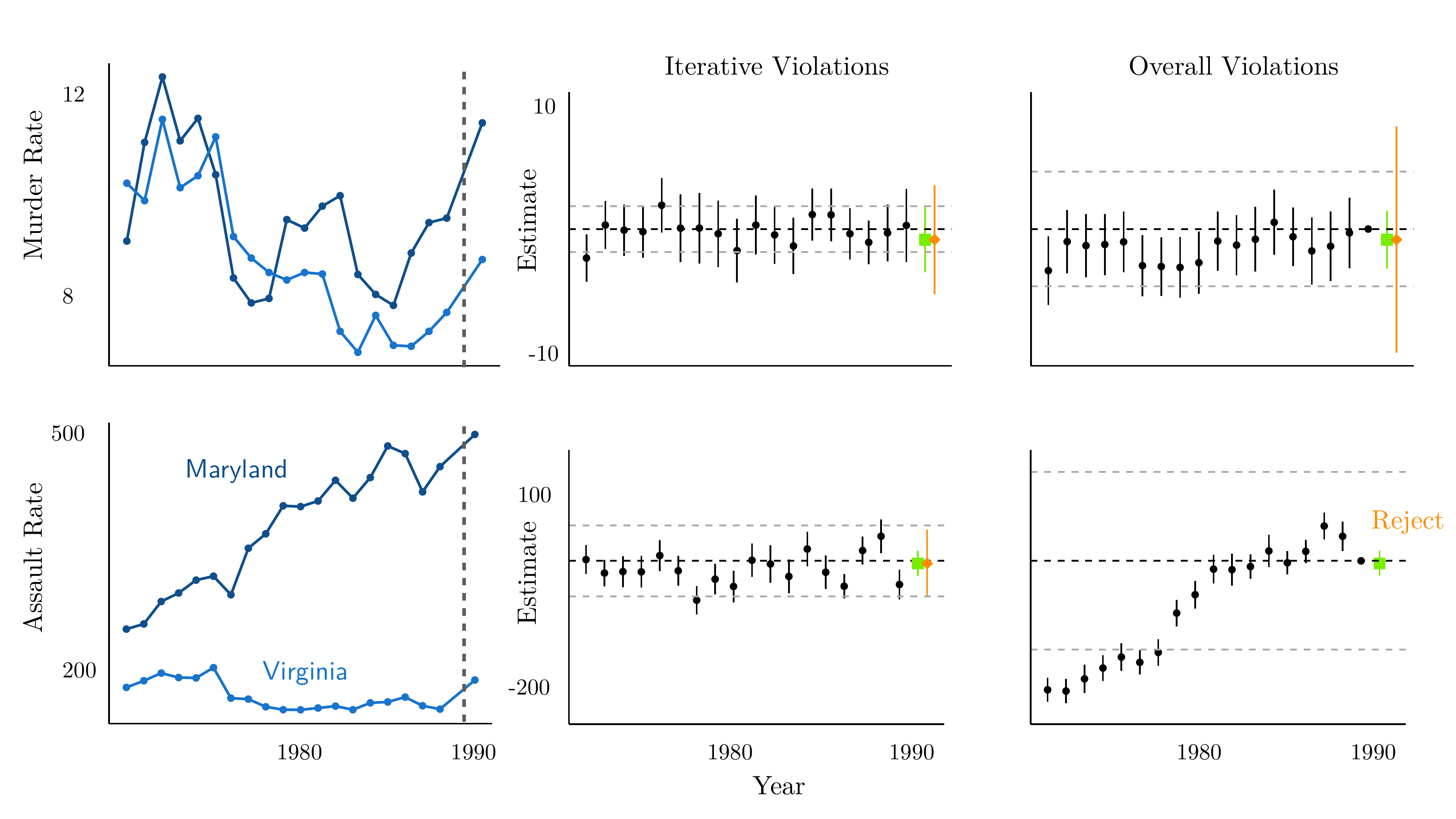}
	\caption{(Top) Murder- and (Bottom) assault state-level rates for Virginia and Maryland \citep{Aneja_Donohue_Zhang_2011}. Event-study plot based on iterative  (Middle) and overall violations (Right).
    The gray, dashed lines indicate the thresholds $M$. We compare the classical DID inference for 1990 (green) with our confidence intervals based on the extrapolation assumption (orange). For assault rates our preliminary test based on the overall violations rejects the extrapolation assumption, and we do not report any inference.}
	\label{fig:crime_rate}
\end{figure}

In this section, we illustrate the proposed methods in a semi-synthetic context.
\citet{manski_how_2018} use a crime rate dataset assembled by \citet{Aneja_Donohue_Zhang_2011} to estimate the effect of right-to-carry (RTC) laws on crimes in Virginia after such legislation was passed in 1989.
They employ a partial identification-based DID approach using Maryland as the control group.
We use this dataset to illustrate the differences that can occur when using the proposed method with iterative violations $\iterviot$ and overall violations $\overvio{t}$, defined in Section~\ref{sec:causal-estimands-and-setup}.

The dataset assembled by \citet{Aneja_Donohue_Zhang_2011} contains crime data aggregated at the state-level.
If counties are viewed as the units of interest, then the crime dataset contains only group means rather than individual observations.
This is a deliberate choice, as county-level crime data have been widely criticized as being unreliable \citep{Maltz_2002,Aneja_Donohue_Zhang_2011}.
In light of this fact, \citet{manski_how_2018} advocate for point estimation and forgo any inferential procedure in their analysis.
For the purposes of our illustration, we create synthetic county-level data based on the state means so that we may illustrate the inferential methods.

We create county-level data using a repeated cross-sectional approach.
Independently for each time $t$ and state $s$, we draw residuals $\widetilde{\epsilon}_{1,t} \dots \widetilde{\epsilon}_{n_s,t} \sim \mathcal{N}(0,\sigma^2)$ and then center them to have zero sample mean, i.e. $\epsilon_{i,t} = \widetilde{\epsilon}_{i,t} - c_t$ where $c_t = n_s^{-1} \sum_{i=1}^{n_s} \widetilde{\epsilon}_{i,t}$.
The observed outcome for county $i$ is set as $Y_{i,t} = \bar{Y}_{t} + \epsilon_{i,t}$, where $\bar{Y}_t$ is the state-level aggregate reported in the dataset.
We use $n_s = 15$ and $n_s = 50$ counties for Maryland and Virginia, respectively, and we set $\sigma = 3$ for murder rates and $\sigma=30$ for assault rates.
We refer to this set-up as semi-synthetic because the point estimator, preliminary test, and bias term in the intervals depend only on the actual group means $\bar{Y}_t$ whereas it is only the fluctuation term $f(\level, \estcov)$ in the confidence interval which depends on the synthetic county-level data.

We then calculate the following estimators 
\begin{align*}
\hat \tau_{DD}^{(1990)} &= \frac{1}{50} \paren[\Big]{
	 \sum_{i = 1}^{50} Y_{i,V}^{(1990)} -\sum_{i = 1}^{50}Y_{i,V}^{(1988)}}
     - \frac{1}{15}  \paren[\Big]{
	 \sum_{i=1}^{15} Y_{i,M}^{(1990)} -\sum_{i=1}^{15}Y_{i,M}^{(1988)}
     } \enspace,
\\
\estiterviot &= 
\frac{1}{50} \paren[\Big]{
	 \sum_{i = 1}^{50} Y_{i,V}^{(t)} -\sum_{i = 1}^{50}Y_{i,V}^{(t-1)}}
     - \frac{1}{15}  \paren[\Big]{
	 \sum_{i=1}^{15} Y_{i,M}^{(t)} -\sum_{i=1}^{15}Y_{i,M}^{(t-1)}} \enspace \text{for} \enspace t\leq 1988 \quadand \\
\estovervio{t} &= 
\frac{1}{50} \paren[\Big]{
	 \sum_{i = 1}^{50} Y_{i,V}^{(t)} -\sum_{i = 1}^{50}Y_{i,V}^{(1988)}}
     - \frac{1}{15}  \paren[\Big]{
	 \sum_{i=1}^{15} Y_{i,M}^{(t)} -\sum_{i=1}^{15}Y_{i,M}^{(1988)}} \enspace \text{for} \enspace t\leq 1988 
\enspace.
\end{align*}

Figure \ref{fig:crime_rate} shows the original state-level murder and assault rates (left) together with event-study plots based on the iterative $\estiterviot$ (middle) and overall violations $\estovervio{t}$ (right).
Murder rates and assault rates show different degrees of violation of parallel trends in the pre-treatment period when comparing Virginia with Maryland. 
A visual inspection shows that while the murder rates of Maryland may plausibly be used to extrapolate to the counterfactual behavior of the rates in Virginia, assault rates in both states evolve very dissimilarly -- making extrapolation questionable.

Using this illustration, we want to compare how our preliminary tests and confidence intervals change based on whether overall violations $\estovervio{t}$ or the iterative violations $\estiterviot$ are used.
For the overall violations $\overvio{t}$, we set extrapolation parameters as $p = \infty$ and $M$ equal to $1/2$ the average outcome.
For iterative violations, we set $p = 1$ and $M$ equal to $1/5$ the average outcome.
The first setting reflects a belief that extrapolation is possible only when all previous overall violations are of a medium magnitude whereas the second setting posits that extrapolation is possible when the average iterative violations are of a smaller magnitude.

Figure \ref{fig:crime_rate} (middle \& right) show the outcome of our inference (orange) based both on the iterative and overall violations.
For murder rates, the preliminary test accepts the extrapolation condition, and we report our confidence interval based on the conditional extrapolation assumption.
The confidence interval based on the overall violations  $\ci^\Delta$ is larger than the one based on the iterative violations $\ci$ because the largest overall violation is substantively bigger than the average iterative violation. 
In the case of assault rates, we report a confidence interval $\ci$ only for the iterative violations: the preliminary test rejects the extrapolation condition for overall violations, and thus we refrain from any further DID analysis.
In this illustration, the event-study plot of overall violations gave a clearer visual illustration of the violation of parallel trends, though this is also due to the trend of their signs.
Based on this example, researchers using iterative violations might want to consider substantially smaller values of $M$ for iterative violations than for overall violations.

	\section{Conclusion} \label{sec:conclusion}

In this paper, we provide a simple preliminary test for the DID research design and accompanying confidence intervals for the ATT. 
We provide conditions under which the preliminary tests are asymptotically consistent and the confidence intervals are valid conditioned on passing the preliminary test.
In this sense, our statistical methods overcome previously identified shortcomings of conventional procedures.
Our method relies on a conditional extrapolation assumption, which we see as a formalization of the assumption which is implicitly held by applied researchers when performing preliminary tests for parallel trends within the DID design.

There are several natural directions for future work.
One direction is to construct preliminary tests based on different formulations of the conditional extrapolation assumption.
For example, it may be practically beneficial to incorporate additional polyhedral restrictions as considered by \citet{rambachan_more_2023}.
Another fruitful direction for future work would be to investigate preliminary testing procedures for alternative DID-type identification strategies including, for example, the higher-order differences for multiple pre-treatment periods \citep{Lee_2016,Mora_Reggio_2019}.
In these cases, the conditional extrapolation assumption itself must be revisited to accommodate the higher-order differences.
Finally, it is natural to ask whether similar preliminary testing procedures will be appropriate in staggered treatment adoption regimes, which have become increasingly popular among economists and political scientists \citep{Sun_Abraham_2021,Goodman-Bacon_2021,Callaway_SantAnna_2021,Athey_Imbens_2022}.
We believe the key challenge will be to devise appropriate conditional extrapolation assumptions which accommodate the staggered regimes.

	
	\printbibliography
	
	\newpage
		
	\appendix
	
	\addcontentsline{toc}{section}{Appendix} 
	\part{Appendix} 
	\parttoc 
	\newpage
	
	\section{Application to Structural Equation Models (TWFE)}
\label{App:Equiv}

In the main paper, we used the potential outcomes framework to frame the difference-in-differences method.
In this section, we show that analogous results on pre-tests will follow under structural models which are also common in econometrics.

Consider the following two-way fixed effects (TWFE) structural model \citep{abadie_semiparametric_2005,angrist_mostly_2009}
\[
Y_{it} = \alpha_i + \lambda_t + \beta D_i \indicator{t = t_0} + \varepsilon_{it}
\enspace.
\]
The parameter $\beta$ can be understood as the average effect of treatment, i.e, $\beta = \E{Y_{it} \lvert D_i = 1} - \E{Y_{it} \lvert D_i = 0}$. Assuming the selection for treatment does not depend on the individual-transitory shocks, that is, $P(D_i \lvert \varepsilon_{it}) = P(D_i)$ for all $i,t$, we have that the treatment effect is identified via difference-in-differences \citep{abadie_semiparametric_2005}
\[
\beta = \paren{\E{Y_{it_0} \lvert D_i = 1} - \E{Y_{it_0-1} \lvert D_i = 1}} -  \paren{\E{Y_{it_0} \lvert D_i = 0} - \E{Y_{it_0-1} \lvert D_i = 0}}. 
\]
It can be shown that the OLS estimator for $\beta$ given by
\[
(\hat \alpha, \hat \lambda, \hat \beta) = \argmin_{\alpha,\lambda,\beta} \sum_{i=1}^n \sum_{t=t_0-1}^{t_0} (Y_{it} - \paren{\alpha_i + \lambda_t + \beta D_i \indicator{t = t_0}})^2
\]
is algebraically equivalent to the non-parametric sample mean estimator \citep[Supp. C.1.2]{egami_using_2023}
\[
\hat \beta = \estdd{t_0} = \frac{\sum_{i:D_i =1 } Y_{it_0} - Y_{it_0-1}}{n_T} - \frac{\sum_{i:D_i =0} Y_{it_0} - Y_{it_0-1}}{n_C},
\]
 where $n_T = \sum_{i: D_i =1} 1$ and $n_C = \sum_{i: D_i =0}1$.
 A similar equivalence can be established for the TWFE model in the case of repeated cross-sections \citep[Supp. C.1.1]{egami_using_2023}.

If there are $\Tpost$ post-treatment time-periods, lag-terms can be included to allow for treatment effects for time-periods succeeding $t_0$:
\[
Y_{it} = \alpha_i + \lambda_t + \sum_{s=t_0}^{T}\beta_s D_i \indicator{t = s} + \varepsilon_{it}.
\]

 When pre-treatment periods are available, researchers commonly include lead coefficients in their regression to check whether the treatment and control group evolve in parallel before treatment is administered \citep{Autor_03,angrist_mostly_2009,egami_using_2023}. This leads to the following extension of the TWFE structural equation model
 \[
 Y_{it} = \alpha_i + \lambda_t + \sum_{s=t_0}^{T} \beta_s D_i \indicator{t = s} + \sum_{s=1}^{t_0-2}\delta_s D_i \indicator{t = s} + \varepsilon_{it},
 \]
 where $\delta_{t_0-1} = 0$ is set to zero for normalization. These lead coefficients 
  \[
  \delta_{s} = \paren{\E{Y_{is} \lvert D_i = 1} - \E{Y_{it_0-1} \lvert D_i = 1}} -  \paren{\E{Y_{is} \lvert D_i = 0} - \E{Y_{it_0-1} \lvert D_i = 0}}
  \]
  can be understood as the structural equation model equivalent to the overall violations of parallel trends $\overvio{s}$ defined in Section \mainref{sec:prelim}.
 In practice, estimates of $\delta_{s}$ are routinely plotted in event-study plots to check for violations of parallel trends in the pre-treatment period \citep{Autor_03,gallagher_learning_2014,fitzpatrick_early_2014}.  We can also define $r_s = \delta_s - \delta_{s-1}$ as the equivalent to the iterative violation of parallel trends $\itervio{s}$ in our main discussion. 
 
 Defining $\popfunc = \paren{r_2,\dots, r_{t_0-1},\beta_{t_0},\dots,\beta_{T}}$ and its estimated equivalent $\estpopfunc = \paren{\hat r_2,\dots, \hat r_{t_0-1},\hat \beta_{t_0},\dots,\hat \beta_{T}}$, our discussion in the main part of this paper immediately applies to the structural equation model perspective on DID. 

\section{Extrapolation Based on the Overall Violations $\overviot$}
\label{App:AbsoluteViolations}

In the main paper, we defined the severity measure of the parallel trends based on the iterative violations.
In this section, we show that everything can be equivalently defined in terms of the overall violations. 

As discussed in the main paper, the overall- and iterative violations of parallel trends are related via
\begin{align}\nonumber
\overviot &= \sum_{s=t_0}^t \itervio{s} \hspace{1.2cm} (t \geq t_0), \\ \nonumber
\overviot &= -\sum_{s=t+1}^{t_0-1} \itervio{s} \hspace{0.5cm} (t \leq  t_0 -1) 
\quad \text{and} \quad \\\nonumber
 \iterviot &= \overvio{t} - \overvio{t-1} \enspace,
\end{align}
and it is straightforward to derive results based on the overall violations $\overviot$ equivalent to those presented in the main text for the iterative violations $\iterviot$.

When working with the overall violations, we define the functional and associated estimator
\[
\popfunc^{\Delta} = \paren[\Big]{ \overvio{1} , \dots \overvio{t_0-2}, \dd{t_0}, \dots \dd{T} }
\quadand
\estpopfunc^{\Delta} = \paren[\Big]{ \estovervio{1},  \dots \estovervio{t_0-2}, \estdd{t_0}, \dots \estdd{T} }
\enspace.
\]
By the Cramér-Wold device and continuous mapping, Assumption~\mainref{a:uniform_asymptotic_normality} and Assumption~\mainref{a:unifConsistency} immediately imply analogue statements for $\popfunc^{\Delta}$. Note that a variance estimator $\estcov$ for $\estpopfunc$ immediately implies a variance estimator $\estcov^{\Delta}$ for $\estpopfunc^{\Delta}$.
\begin{lemma} \label{a:uniform_asymptotic_normality_delta}
    Suppose Assumption \mainref{a:uniform_asymptotic_normality} and \mainref{a:unifConsistency} holds. Then $\popfunc^{\Delta}_n \to \popfunc^{\Delta}$ and $\sqrt{n} \cdot \paren[\big]{ \popfunc^{\Delta}_n - \estpopfunc^{\Delta}_n }
	\xrightarrow{d}
	\mathcal{N}(0,\pfcov^{\Delta})$.
    Moreover, the variance estimator satisfies $\estcov^{\Delta}- \pfcov^{\Delta}\xrightarrow{p} 0$.
\end{lemma}

\subsection{Extrapolation Assumption on the Overall Violations}
We define the severity of overall parallel trends violations
\[
\Spre^{\Delta} = \paren[\Bigg]{ \frac{1}{\Tpre-1} \sum_{t = 2}^{t_0-1} \abs[\big]{ \overviot}^p }^{1/p}
\quadand
\Spost^{\Delta}= \paren[\Bigg]{ \frac{1}{\Tpost} \sum_{t = t_0}^{T} \abs[\big]{ \overviot }^p }^{1/p}
\enspace.
\]
We can then make a conditional extrapolation assumption based on the overall violations: 
\begin{assumption}[Conditional Extrapolation (Overall Violations)] \label{a:extrp_over}
	If $\Spre^{\Delta} \leq M$, then $\Spost^{\Delta} \leq \Spre^{\Delta}$.
\end{assumption}
The following proposition bounds the difference between the average post-treatment DID and average post-treatment ATT:
\begin{proposition} \label{prop:bias_over}
Suppose Assumption~\suppref{a:extrp_over} (conditional extrapolation based on the overall violations) holds.
If $\Spre^{\Delta}  \leq M$, then the difference between the average post-treatment DID and average post-treatment ATT is bounded as
\[
  \abs[\big]{ \totalatt - \totaldd }
  \leq  \Spre^{\Delta} \enspace.
\]
If $\Spre^{\Delta}  > M$, then $\abs{ \totalatt - \totaldd }$ may be arbitrarily large.
\end{proposition}
\begin{proof}
     By definition, we have
     \[
     \totalatt = 
     \totaldd  -  \frac{1}{\Tpost} \sum_{t=t_0}^{T} \overviot.
     \]
The conditional extrapolation assumption based on overall violations (Assumption~\suppref{a:extrp_over}) implies
   \begin{align}
     \abs[\big]{ \totalatt -   \totaldd }
     &=
   \abs[\Bigg]{  \frac{1}{\Tpost} \sum_{t=t_0}^{T} \overviot }
   \\ 
    \, &\leq \, \Spost^{\Delta} 
    \\ 
    \, &\leq \, \Spre^{\Delta}, 
   \end{align}
where the first inequality follows from Hölder's inequality and the second one from the conditional extrapolation assumption (Assumption~\suppref{a:extrp_over}).
\end{proof}
\subsection{Testing for Overall Violations of Parallel Trends in the Pre-Treatment Period}
It is straightforward to derive equivalent results to those in the main text based on the overall violations $\overviot$. We begin by providing results on testing for overall violations of parallel trends in the pre-treatment period. We define the $s_n$-separated null and alternative hypotheses
\begin{align*}
	H_0^{\Delta}(s_n) &= \setb[\big]{ \textrm{ Sequences } \popdist_n \textrm{ satisfying Assumption~\mainref{a:uniform_asymptotic_normality} and } \Spre^{\Delta} \leq M - s_n } \\
	H_1^{\Delta}(s_n) &= \setb[\big]{  \textrm{ Sequences } \popdist_n \textrm{ satisfying Assumption~\mainref{a:uniform_asymptotic_normality} and } \Spre^{\Delta} \geq M + s_n }
	\enspace,
\end{align*}
and the estimated severity of overall pre-treatment violations
\[
\estSpre^{\Delta} = \paren[\Big]{ \frac{1}{\Tpre-1} \sum_{t=2}^{t_0-1} \abs[\big]{ \hat {\Delta}^{(t)}  }^p }^{1/p} \enspace.
\]
Analogously to our test in Section~\mainref{Sec:Testing}, we define the simple test 
\[
\phi^{\Delta} = \indicator{ \estSpre^{\Delta} \geq M }
\enspace
\]
to check whether the extrapolation condition based on the overall violations (Assumption \suppref{a:extrp_over}) holds.
The following theorem provides a result on the asymptotic consistency of this test:
\begin{theorem} \label{thm:testing_over}
	Under Assumption~\mainref{a:uniform_asymptotic_normality} (CLT), the test $\phi^{\Delta} $ described above is asymptotically consistent for separation $s_n = \littleOmega{n^{-1/2}}$.
\end{theorem}
\begin{proof}
    The proof follows exactly along the lines of the proof for Theorem~\mainref{thm:testing} in Appendix~\suppref{App:ProofTesting}.
\end{proof}
\subsection{Conditionally Valid Inference for Extrapolation based on Overall Violations}
Following along the lines of Section~\mainref{Sec:Inference}, we define the following confidence interval 
\begin{equation} \label{Eq:CI}
       \ci^{\Delta} = 
        \esttotaldd
        \pm 
        \braces[\Big]{
        	\estSpre^{\Delta}
        	+ \frac{ f^{\Delta}(\level, \estcov^{\Delta}) }{ \sqrt{n}}
    	} \enspace
\end{equation}
for inference under the conditional extrapolation assumption on the overall violations. 
The function $f^{\Delta}(\level, \covsig)$ describes the $(1-\alpha)$ critical values of the limiting distribution corresponding to statistical errors in $\esttotaldd$ and $\estSpre^{\Delta}$.
To do so, we consider the multivariate function $\psi : \Reals^{T-1} \to \Reals$ defined as
\[
\psi^{\Delta}(x_1 \dots x_{T-1})  = \abs[\Big]{ \frac{1}{\Tpost} \sum_{t=\Tpre}^{T-1} x_t }
+ \paren[\Big]{ \frac{1}{\Tpre-1} \sum_{t=1}^{\Tpre-1} \abs{ x_t }^p }^{1/p}
\enspace,
\]
which is constructed to reflect the errors in $\esttotaldd$ and $\estSpre^{\Delta}$.
The value $f^{\Delta}(\level, \covsig^{\Delta})$ is then the critical value of $\psi^{\Delta}(Z)$, where $Z \sim \mathcal{N}(0,\estcov^{\Delta})$.

\begin{theorem}
\label{theorem:cond-inf_over}
	Suppose that Assumptions~\mainref{a:uniform_asymptotic_normality},\mainref{a:unifConsistency}, and~\suppref{a:extrp_over} hold.
	Then, the confidence interval $\ci^{\Delta}$ is conditionally valid on the results of the hypothesis test $\phi^\Delta$ under the well-separated null; that is, if $P_n \in H_0^\Delta(s)$ for $s_n = \littleOmega{n^{-1/2}}$, then
	\[
	\liminf_{n \to \infty } \Pr[\big]{ \totalatt \in \ci^\Delta \mid \phi^\Delta = 0 } \geq 1 - \level
	\enspace.
	\]
\end{theorem} 
\begin{proof}
    This theorem is proved analogously to Theorem~\mainref{theorem:cond-inf}.
    See Appendix~\suppref{App:ProofPropCondInf} for details.
\end{proof}

\section{Extension to General Linear Estimands}
\label{App:OtherEstimands}
In this section, we will extend our discussion to estimands that are linear functions of the individual post-treatment ATTs $\attt$. That is, we will consider the weighted average ATT
\[
\linatt = \sum_{t=t_0}^T c_t \attt.
\]
This class of estimands contains the post-treatment average discussed in the main text, any selection or linear contrast between ATTs at individual time-periods, and general weighted averages that could, for example, discount time-periods based on the time passed since treatment was administered.

The estimand $\linatt$ is not identified in the presence of post-treatment violations of parallel trends. We define the DID equivalent
\[
\lindd = \sum_{t=t_0}^T c_t \ddt.
\]
The following proposition bounds the difference between the ATT $\linatt$ and DID $\lindd$ estimand under the conditional extrapolation assumption (Assumption~\mainref{a:extrapolation}).
\begin{proposition}
Suppose conditional extrapolation (Assumption~\mainref{a:extrapolation}) holds.
If $\Spre \leq M$, then the difference between the weighted average post-treatment DID and weighted average post-treatment ATT is bounded as
\[
  \abs[\big]{ \linatt- \lindd }
  \leq \extrapbiaslin \cdot \Spre \enspace,
\]
where $\extrapbiaslin
=
\Tpost^{1/p}
\left(
\sum_{s=t_0}^{T}
\left|
\sum_{t=s}^{T} c_t
\right|^q
\right)^{1/q}$ for $q$ satisfying $\frac{1}{q}+\frac{1}{p} =1$.
If $\Spre > M$, then $\abs{  \linatt- \lindd }$ may be arbitrarily large.
\end{proposition}
\begin{proof}
    The proof is a simple application of the conditional extrapolation assumption and Hölder's inequality. The result follows along the same lines as the proof of Proposition~\mainref{prop:bias}.
\end{proof}
Given this proposition, it is straightforward to extend the inference under the conditional extrapolation assumption developed in Section~\mainref{Sec:Inference} to the weighted average ATT.
In the definition of the confidence interval $\ci$ we replace $\extrapbias$ by $\extrapbiaslin$.
An analogue to Theorem~\mainref{theorem:cond-inf} then follows immediately.

\section{Proofs of Main Results}

\subsection{Technical Lemmas} \label{supp:technical-lemmas}

In this section, we provide the proof of various technical lemmas.
The following lemma shows that the bound in Proposition~\mainref{prop:bias} is sharp.

\begin{lemma}
    \label{lemma:sharp}
    The upper bound in Proposition~\mainref{prop:bias} is sharp.
\end{lemma}
\begin{proof}
    From the proof of Proposition~\mainref{prop:bias}, we have 
     \begin{align}
     \bigg\lvert     \totalatt -   \totaldd \bigg \rvert 
     \, &= \,  
   \bigg \lvert  \frac{1}{T_{post}} \sum_{t=t_0}^{T} \overviot \bigg\rvert 
   \\\nonumber 
   \, &= \,  
     \bigg \lvert  \frac{1}{T_{post}} \sum_{t=1}^{T_{post}} (T_{post} - t+1)\itervio{t_0-1+t} \bigg\rvert 
      \\\nonumber
      \, &\leq \,  
     \bigg( \frac{1}{T_{post}} \sum_{t=1}^{T_{post}} (T_{post} - t+1)^q\bigg)^{1/q} \bigg( \frac{1}{T_{post}} \sum_{t=1}^{T_{post}} \lvert\itervio{t_0-1+t}\rvert^p\bigg)^{1/p}
      \\\nonumber
      \, &\leq \,  
     \bigg( \frac{1}{T_{post}} \sum_{t=1}^{T_{post}} (T_{post} - t+1)^q\bigg)^{1/q} \Spre
       \\\nonumber
      \, &\equiv \,  
     \extrapbias \cdot \Spre,
\end{align}
where $q$ and $p$ are conjugate (i.e., $1/q + 1/p = 1$).The first inequality is an application of Hölder's inequality.
The second inequality follows from conditional extrapolation (Assumption \mainref{a:extrapolation}.)
For the first inequality to be tight $\sgn{\itervio{t}}$ needs to be positive for all $t \in \setb{t_0,\dots,T}$.
Additionally, for Hölder's inequality to be tight, we need $\lvert \itervio{t_0-1+t}\rvert^p \propto (\Tpost -t +1 )^q$ for all $t \in \setb{1,\dots,\Tpost}$. The last inequality is tight if $\Spre = \Spost$. 
Hence, the bound is sharp, that is, it is attained if $\Spre = \Spost$, all  $\itervio{t}$ in the post-treatment period are positive, and $\lvert \itervio{t_0-1+t}\rvert^p \propto (\Tpost -t +1 )^q$ for all $t \in \setb{1,\dots,\Tpost}$.
\end{proof}

The following lemma shows that the quantile function used in the confidence intervals is continuous.
Recall that we define the multivariate function $\psi : \Reals^{T-1} \to \Reals$ as
\[
\psi(x_1 \dots x_{T-1})  = \abs[\Big]{ \frac{1}{\Tpost} \sum_{t=t_0}^{T-1} x_t }
+ \extrapbias \cdot \paren[\Big]{ \frac{1}{\Tpre-1} \sum_{t=1}^{t_0-1} \abs{ x_t }^p }^{1/p}
\enspace.
\]

\begin{lemma}
\label{lemma:quantile_function}
	If $Z \sim \mathcal{N}(0,\covsig)$ with $\covsig$ non-singular, then
	the quantile function $F^{-1}_{\covsig}$ of $\psi(Z)$ is continuous.
\end{lemma}
\begin{proof}
Let $F_{\psi}(y) = P(\psi(Z) \leq y)$. The quantile function is then given by
$F^{-1}(q,\covsig) = \inf\{y \in \Reals: F_{\psi}(y) \geq q\}$. It is enough to show
that $F_{\psi}$ is continuous and strictly increasing on the support of $\psi(Z)$.
Because $\psi(\cdot)$ is continuous (and $\Reals^{T-1}$ is connected), $\psi(\cdot)$
has a connected image. Consequently, $F_{\psi}$ is strictly increasing on its support.
Hence, it is enough to verify that $P(\psi(Z) = y) = 0 $ for all $y \in \Reals$.

We show that $Y= \psi(Z)$ admits a density. Write
\[
    s(x)=\frac{1}{\Tpost}\sum_{t=\Tpre}^{T-1}x_t
\]
and
\[
    r_p(x)=
    \begin{cases}
    \left(\frac{1}{\Tpre-1}\sum_{t=1}^{\Tpre-1}|x_t|^p\right)^{1/p},
        & 1\leq p<\infty,\\[1.2ex]
    \max_{1\leq t\leq \Tpre-1}|x_t|,
        & p=\infty.
    \end{cases}
\]
Then $\psi(x)=|s(x)|+\kappa r_p(x)$. Note that $\psi(x)$ is Lipschitz and hence
differentiable a.e. by Rademacher's theorem.

We claim that $\norm{\nabla \psi(x)}>0$ at every point where $\psi$ is differentiable.
Let $v\in\Reals^{T-1}$ be the vector with entries equal to zero in the first
$\Tpre-1$ coordinates and entries equal to one in the remaining $\Tpost$ coordinates.
Since $r_p$ depends only on the first $\Tpre-1$ coordinates,
\[
    \psi(x+hv)-\psi(x)=|s(x)+h|-|s(x)|.
\]
If $s(x)=0$, then the right and left derivatives of the previous display at $h=0$
are different. Thus $\psi$ is not differentiable at $x$. Hence, at every point where
$\psi$ is differentiable, we have $s(x)\neq 0$. At such a point,
\[
    \frac{\partial \psi}{\partial x_t}(x)
    =
    \frac{\sgn{s(x)}}{\Tpost},
    \qquad t=\Tpre,\dots,T-1.
\]
Therefore,
\[
    \norm{\nabla \psi(x)}
    \geq
    \frac{1}{\sqrt{\Tpost}}
\]
at every point where $\psi$ is differentiable. In particular,
\[
    \norm{\nabla \psi(x)} > 0 \ \ \text{a.e.}
\]
and $1/\norm{\nabla\psi(x)}\leq \sqrt{\Tpost}$ a.e.

Now fix an arbitrary Borel set $A$ and define
\[
     h_A(z) =
     \frac{\phi(z)}{\norm{\nabla \psi(z)}}\indicator{\psi(z) \in A},
\]
where $\phi(z)$ is the normal density of $Z$ on $\Reals^{T-1}$. We set $h_A(z)=0$
on the set where $\psi$ is not differentiable. By the previous display,
\[
    0\leq h_A(z)\leq \sqrt{\Tpost}\,\phi(z),
\]
and hence $h_A$ is integrable.

Applying the coarea formula \citep[Thm 3.11]{Evans_Gariepy_2015}, we have
\begin{equation} \label{eq:coarea}
   \int_{\Reals^{T-1}} h_A(z) \norm{ \nabla  \psi(z) } dz
   =
   \int_\Reals \paren[\Big]{ \int_{\psi^{-1}(y)} h_A(z) d\mathcal{H}^{T-2}(z) } dy,
\end{equation}
with $d\mathcal{H}^{T-2}$ being the $T-2$-dimensional Hausdorff measure on the level
set $\psi^{-1}(y)$.

The left-hand side of the previous display is
\[
    \int_{\Reals^{T-1}}
    \phi(z) \cdot \indicator{\psi(z) \in A} dz
    =
    P(\psi(Z) \in A).
\]
For the right-hand side of Equation~\eqref{eq:coarea}, note that
$\indicator{\psi(z)\in A}=\indicator{y\in A}$ on the level set $\psi^{-1}(y)$.
Hence,
\[
   P(\psi(Z) \in A)
   =
   \int_A
   \bigg(
        \int_{\psi^{-1}(y)}
        \frac{\phi(z)}{\norm{\nabla \psi(z)}} d\mathcal{H}^{T-2}(z)
   \bigg) dy,
\]
where the integrand is set to zero at points where $\psi$ is not differentiable.
Hence, $\psi(Z)$ admits a density:
\[
    f_\psi(y)
    =
    \int_{\psi^{-1}(y)}
    \frac{\phi(z)}{\norm{\nabla \psi(z)}} d\mathcal{H}^{T-2}(z).
\]
 This concludes the proof.
\end{proof}
\subsection{Bound on the Bias (Proposition~\mainref{prop:bias})}

In this section, we prove the following proposition, which bounds the bias of the difference-in-differences estimator for the average treatment effect on the treated.

\begin{refproposition}{\mainref{prop:bias}}
	\biasbound
\end{refproposition}

\begin{proof}
    By definition, we have
\[
    \totalatt = \totaldd  -  \frac{1}{T_{post}} \sum_{t=t_0}^{T} \overviot \enspace.
\]
The conditional extrapolation assumption (Assumption~\mainref{a:extrapolation}) implies
   \begin{align}
     \bigg\lvert     \totalatt -   \totaldd \bigg \rvert 
     \, &= \,  
   \bigg \lvert  \frac{1}{T_{post}} \sum_{t=t_0}^{T} \overviot \bigg\rvert 
       \\\nonumber
     \, &= \,  
     \bigg \lvert  \frac{1}{T_{post}} \sum_{t=t_0}^{T} \sum_{s=t_0}^t \itervio{s} \bigg\rvert 
      \\\nonumber
     \, &= \,  
     \bigg \lvert  \frac{1}{T_{post}} \sum_{t=1}^{T_{post}} (T_{post} - t+1)\itervio{t_0-1+t} \bigg\rvert 
       \\\nonumber
      \, &\leq \,  
     \bigg( \frac{1}{T_{post}} \sum_{t=1}^{T_{post}} (T_{post} - t+1)^q\bigg)^{1/q} \bigg( \frac{1}{T_{post}} \sum_{t=t_0}^{T} \lvert\iterviot\rvert^p\bigg)^{1/p}
      \\\nonumber
      \, &\leq \,  
     \bigg( \frac{1}{T_{post}} \sum_{t=1}^{T_{post}} (T_{post} - t+1)^q\bigg)^{1/q} \Spre
       \\\nonumber
      \, &\equiv \,  
     \extrapbias \cdot \Spre
\end{align}
where the second to last inequality follows from Hölder's inequality by choosing $q \in [1,\infty]$ such that $\frac{1}{q}+\frac{1}{p} =1$.
The last inequality follows from the conditional extrapolation assumption.
Note that by reindexing, we can write $\extrapbias =  \paren[\bigg]{\frac{1}{T_{post}} \sum_{t=1}^{T_{post}}t^q}^{1/q}$.
    
\end{proof}

\subsection{Asymptotically Consistent Test (Theorem~\mainref{thm:testing})}
\label{App:ProofTesting}

In this section, we prove Theorem~\mainref{thm:testing}, which establishes that the test of the extrapolation condition is asymptotically consistent.

\begin{reftheorem}{\mainref{thm:testing}}
	\testingthm
\end{reftheorem}

\begin{proof}
Consider a sequence $\popdist_n^{(0)}  \in H_0(s_n)$. Then for arbitrary $n \in \mathbb{N}$, we have
\begin{align}
  P(\estSpre  \geq M) 
  \, & =\,   
  P(\estSpre - \Spre\geq M - \Spre ) 
  \\[1em]\nonumber
  \, &\leq \, 
  P(\estSpre- \Spre\geq s_n ) 
 \\[1em]\nonumber
  \, &\leq \, 
  P(\big\lvert\estSpre - \Spre \big \rvert \geq s_n ) 
   \\[1em]\nonumber
  \, &\leq \, 
  P(\frac{\big\lvert\estSpre - \Spre\big\rvert}{s_n}  \geq 1 ).
\end{align}
Define the vector $R_n = \paren{ \itervio{2} , \dots \itervio{\Tpre}}$ and its estimated counterpart $
\hat R_n = \paren{ \estitervio{2},  \dots \estitervio{\Tpre}}$, which we index by $n$ to indicate their sample size dependence.
By Assumption~\mainref{a:uniform_asymptotic_normality}, we have that $\sqrt{n}(\hat R_n - R_n) \darrow N(0,\tilde \covsig)$ along the considered sequence $\popdist_n^{(0)}$
for some covariance matrix $\tilde \covsig$. By Prokhorov's theorem \citep[Thm. 2.4]{vaart_asymptotic_1998}, this implies that $\{\sqrt{n}(\hat R_n - R_n)\}_{n \in \mathbb{N}}$ is uniformly tight, that is, $\sqrt{n}(\hat R_n - R_n) =\bigOp{1}$. Hence, $(\hat R_n - R_n) = \bigOp{1/\sqrt{n}}$.
Now note that $\lVert 1/(\Tpre-1) (\hat R_n - R_n) \rVert_p\leq \lVert 1/(\Tpre-1) (\hat R_n - R_n) \rVert_1$ for all $p\geq 1$ and hence $\lVert 1/(\Tpre-1) (\hat R_n - R_n) \rVert_p=\bigOp{1/\sqrt{n}}$.
Lastly, we have $\big\lvert\estSpre-  \Spre \big\rvert \leq (\Tpre -1)^{-1/p} \lVert (\hat R_n - R_n) \rVert_p=\bigOp{1/\sqrt{n}} $ by the reverse-triangle inequality.

With $s_n = \littleOmega{n^{-1/2}}$, we have $\frac{\big\lvert\estSpre - \Spre\big\rvert}{s_n} = \littleOp{1}$ and
hence
     \begin{align}
         \lim_{n\rightarrow \infty}  P(\estSpre  \geq M)  =\lim_{n\rightarrow \infty} \Esub{X \sim \popdist_n^{(0)}}{\phi(X)}=  0
     \end{align}
    along the considered sequence $\popdist_n^{(0)}$.

    The second half of the proposition follows similarly.
Consider a sequence $\popdist_n^{(1)} \in H_1^{(n)}(s_n)$. Then for arbitrary $n \in \mathbb{N}$, we have
\begin{align}
  P(\estSpre  \leq M) 
  \, & =\,   
  P(\estSpre -\Spre   \leq M -\Spre  ) 
  \\[1em]\nonumber
  \, &\leq \, 
  P(\estSpre -\Spre   \leq - s_n ) 
 \\[1em]\nonumber
  \, &\leq \, 
  P(\big\lvert\Spre   - \estSpre\big \rvert \geq s_n ) 
   \\[1em]\nonumber
  \, &= \, 
  P(\frac{\big\lvert\estSpre - \Spre\big\rvert}{s_n}  \geq 1 )
\end{align}
By the preceding asymptotic argument, we have
\begin{align}
     \lim_{n\rightarrow \infty}  P(\estSpre  \leq M)  =\lim_{n\rightarrow \infty} \Esub{X \sim \popdist_n^{(1)}}{1- \phi(x)}=  0
\end{align}
along the considered sequence $\popdist_n^{(1)}$. This concludes the proof.
\end{proof}

\subsection{Conditionally Valid Inference (Theorem~\mainref{theorem:cond-inf})}
\label{App:ProofPropCondInf}

In this section, we prove that the proposed confidence intervals are conditionally valid when the extrapolation condition is true and the separation is sufficiently large.

\begin{reftheorem}{\mainref{theorem:cond-inf}}
	\conditionalcoverage
\end{reftheorem}
\begin{proof}
We will first show that $\ci$ is asymptotically valid under Assumption \mainref{a:uniform_asymptotic_normality} and then show that proper coverage is guaranteed even when conditioned on the test of Theorem \mainref{thm:testing}.

We begin by defining the vector $R_n = \paren{ \itervio{2} , \dots \itervio{\Tpre}}$ and its estimated counterpart $\hat R_n = \paren{ \estitervio{2},  \dots \estitervio{\Tpre}}$, which we index by $n$ because of their sample size dependence. Now consider 
\begin{align}
    &P\bigg(\lvert \totalatt - \esttotaldd \rvert
    \, \geq\,  
     \extrapbias \estSpre + \hcrvn \frac{1}{\sqrt{n}}\bigg ) 
    \\\nonumber
    \, &= \, 
     P\left( \lvert \totalatt - \esttotaldd \rvert -  \extrapbias \estSpre\geq   \hcrvn \frac{1}{\sqrt{n}}\right )  
      \\\nonumber
    \, &= \, 
     P\left(\lvert \totalatt - \esttotaldd \rvert -\extrapbias \Spre +\extrapbias \Spre -\extrapbias \estSpre\geq   \hcrvn \frac{1}{\sqrt{n}}\right )  
     \\\nonumber
    \, &\leq \, 
     P\left(\lvert \totalatt - \esttotaldd \rvert - \lvert \totalatt- \totaldd \rvert  +\extrapbias  ( \Spre-\estSpre ) \geq   \hcrvn \frac{1}{\sqrt{n}}\right )  
        \\\nonumber
    \, &\leq \, 
     P\left(\bigg\lvert \lvert \totalatt - \esttotaldd \rvert - \lvert \totalatt- \totaldd \rvert \bigg \rvert +\extrapbias  \bigg\lvert \Spre-\estSpre \bigg\rvert \geq   \hcrvn \frac{1}{\sqrt{n}}\right )  
        \\\nonumber
    \, &\leq \, 
     P\left( \lvert  \totaldd- \esttotaldd \rvert +\extrapbias  (\Tpre-1)^{-1/p} \rVert R - \hat R\lVert_{p}  \geq   \hcrvn \frac{1}{\sqrt{n}}\right )
         \\\nonumber
    \, &= \, 
     P\left(\sqrt{n}\bigg( \lvert  \totaldd- \esttotaldd  \rvert +\extrapbias  (\Tpre-1)^{-1/p} \rVert R - \hat R\lVert_{p} \bigg) \geq   \hcrvn \right ),  
\end{align}
where the last inequality follows from applying the reverse triangle inequality to both terms.
\sloppy
We remind ourselves of the definitions \[
\popfuncn = \paren[\Big]{ \itervio{2} , \dots \itervio{t_0-1}, \dd{t_0}, \dots \dd{T} }
\quadand
\estpopfuncn = \paren[\Big]{ \estitervio{2},  \dots \estitervio{t_0-1}, \estdd{t_0}, \dots \estdd{T} }\enspace.\]
We then define the random variable 
\begin{align}\nonumber
    Y_n\, &= \,\sqrt{n} \bigg (\lvert  \totaldd - \esttotaldd \rvert +\extrapbias  (\Tpre-1)^{-1/p} \rVert R - \hat R\lVert_{p}  \bigg) \equiv \psi\big(\sqrt{n}(\popfuncn - \estpopfuncn)\big).
\end{align}
Notice that $\psi$ is a continuous function. 
Under Assumption \mainref{a:uniform_asymptotic_normality} and by the continuous mapping theorem, we have
\begin{align}\nonumber
    \psi\big(\sqrt{n}(\popfuncn - \estpopfuncn)\big) \darrow \psi(Z),
\end{align}
where $Z \sim \mathcal{N}(0,\covsig_\popfunc)$.

Given Lemma \suppref{lemma:quantile_function} and Assumption \mainref{a:unifConsistency}, we have $\hcrvn \parrow \crv$ by the continuous mapping theorem.
We then have 
\begin{align}\nonumber
     P\left(\sqrt{n}\bigg( \lvert  \totaldd - \esttotaldd \rvert +\extrapbias  (\Tpre-1)^{-1/p} \rVert R - \hat R\lVert_{p} \bigg) \geq   \hcrvn \right ) \rightarrow P(\psi(Z) \geq \crv) = \alpha.
\end{align}
This shows that $\ci$ is asymptotically valid. 

Now consider a sequence $\popdist_n^{(0)} \in H_0^{(n)}(s_n)$ and let $s_n = \littleOmega{n^{-1/2}}$. For any $n \in \mathbb{N}$, we have by the law of total probability
    \begin{align}
        P(\totalatt \in \ci) 
        \, &= \,
        P\left(\totalatt \in \ci |\phi = 0\right) P(\phi = 0) 
        \\\nonumber
        \, & + \, 
         P\left(\totalatt \in \ci |\phi = 1\right) P(\phi = 1) 
        \\\nonumber
        \, & \leq  \,  P\left(\totalatt \in \ci |\phi = 0\right) P(\phi = 0) 
         \\\nonumber
        \, & + \, 
        P(\phi = 1).
    \end{align}
By Proposition \mainref{thm:testing}, we have $\lim_{n\rightarrow\infty} P(\phi = 1 )= 0$ and $\lim_{n\rightarrow\infty} P(\phi = 0) = 1$ along the considered sequence $\popdist_n^{(0)}$. Consequently, we have
\begin{align}
    1- \alpha 
    \, &\leq \, 
    \liminf_{n\rightarrow\infty} P\left(\totalatt \in \ci |\phi = 0\right)  
    \\\nonumber
    \, &\leq \, \liminf_{n\rightarrow\infty} \bigg( P\left(\totalatt \in \ci |\phi = 0\right) P(\phi = 0) 
    +
    P(\phi = 1) \bigg)
         \\\nonumber
        \, & = \, 
        \liminf_{n\rightarrow\infty}P\left(\totalatt \in \ci |\phi = 0\right).
\end{align}
This concludes the proof.
\end{proof}
\subsubsection{Calculation of the critical values $f(\level, \estcov)$}
Recall the multivariate function $\psi : \Reals^{T-1} \to \Reals$:
\[
\psi(x_1 \dots x_{T-1})  = \abs[\Big]{ \frac{1}{\Tpost} \sum_{t=\Tpre}^{T-1} x_t }
+ \extrapbias \cdot \paren[\Big]{ \frac{1}{\Tpre-1} \sum_{t=1}^{\Tpre-1} \abs{ x_t }^p }^{1/p}
\enspace.
\]
The critical value $f(\level, \estcov)$ is defined as satisfying $\Pr{ \psi(Z) \geq f(\level, \estcov) } = \alpha$, where $Z \sim \mathcal{N}(0,\estcov)$. While there is no closed form solution for the quantile function of $\psi(Z)$ in general, we can obtain the critical value numerically. 
We do so via Monte-Carlo approximation by drawing $Z^{(s)} \overset{iid}{\sim} \mathcal{N}(0,\estcov)$ for $s \in \setb{1,\dots,S}$ with $S=5000$.
A Monte-Carlo approximation of the critical value is then given by
\[
f(\level, \estcov) \approx \max \setb[\bigg]{ c_f \in \mathbb{R}_+ : \frac{1}{S}\sum_{s=1}^S \indicator{\psi(Z^{s}) \geq c_f} \geq \level } \enspace .
\]

\subsection{Probability of Valid Reporting (Theorem~\mainref{theorem:uncond-inf})}

In this section, we prove that the probability of valid reporting for the proposed test and confidence interval is of at least nominal coverage asymptotically when the extrapolation condition is true and the separation is sufficiently large.

\begin{reftheorem}{\mainref{theorem:uncond-inf}}
	\unconditionalcoverage
\end{reftheorem}
\begin{proof}
	We have that 
    \[
\liminf_{n\rightarrow\infty} \Pr{ \totalatt \in \ci \text{ and } \phi = 0 } \geq \liminf_{n\rightarrow\infty} \Pr{\totalatt \in \ci \lvert \phi =0} \cdot \liminf_{n\rightarrow\infty} \Pr{\phi = 0} \enspace .
\]
Under the assumptions of the theorem, we have that 
\[
\liminf_{n\rightarrow\infty} \Pr{\totalatt \in \ci \lvert \phi =0} \geq 1- \alpha
\]
by Theorem~\mainref{theorem:cond-inf}. Additionally, we have 
\[
\lim_{n\rightarrow\infty} \Pr{\phi = 0} =1 
\]
by Theorem~\mainref{thm:testing}. Hence
\[\liminf_{n\rightarrow\infty} \Pr{ \totalatt \in \ci \text{ and } \phi = 0 } \geq 1- \alpha \enspace . \]

\end{proof}

\section{Additional Details of Illustrative Re-analysis} \label{app:empirical-example}

In this section, we provide additional details of our re-analysis in Section~\mainref{sec:illustration}.
We use the \citet{Malesky_Nguyen_Tran_2014} difference-in-differences study of the effect of recentralization on public infrastructure in Vietnam.

Following heated debate about the government's proposal to dissolve District People's Councils (DPC) -- a governance structure on an intermediate scale between provinces and communes -- Vietnam's leadership decided to run a pilot program to empirically determine the effects of recentralization.
Treatment, that is, recentralization, was carefully stratified based on region, rurality, and socioeconomic and public administration performance, to mitigate selection bias. In 2009, the DPCs of ten such selected provinces were dissolved. 
\citet{Malesky_Nguyen_Tran_2014} use data from the Vietnam Household Living Standard Surveys in 2008 and 2010 for a difference-in-differences analysis to determine the effect of the dissolution of DPCs on a range of public infrastructure outcomes.
While the stratified treatment assignment lends some ex ante credibility to the parallel trends assumption, it is unlikely that all outcomes of interest would have evolved in parallel in the treatment and control districts absent of recentralization. \citet{Malesky_Nguyen_Tran_2014} use an available pre-treatment period, data from the year 2006, to test for violations of parallel trends. 
While for some outcomes parallel trends appear credible, for many only weaker forms of extrapolation may be justifiable \citep{egami_using_2023}.
For a minority of the outcomes, any extrapolation from the control to the counterfactual trend of the treated districts seems dubious. 
Following \citet{egami_using_2023}, we focus on three paradigmatic outcomes: the existence of programs investing into education or culture (\emph{Educational and Cultural Program}), whether tap water is the main source of drinking water (\emph{Tap Water}), and the existence of an agriculture extension center (\emph{Agricultural Center}).  See the first row of Figure~\mainref{fig:malesky} for the evolution of the mean for these outcomes under treatment and control. From left to right, the parallel trends assumption becomes less plausible, culminating in the right panel for which the possibility of extrapolation seems questionable.

For all three outcomes, we calculate a Horvitz-Thompson estimate of the pre-treatment violation of parallel trends $\estitervio{0}$ and the treatment effect $\esttotaldd$ adjusting for four independent variables deemed necessary for valid extrapolation \citep{Malesky_Nguyen_Tran_2014}.
To take into account that the sampled communes are clustered on the district-level, we calculate an estimate of the covariance matrix $\estcov$ via block-bootstrap \citep{egami_using_2023}.
We determine a threshold $M$ for acceptable violations of pretreatment violations of parallel trends based on the scale of the outcomes.
More specifically, we set $M$ to be half the overall level of the respective outcome measure.
For violations more severe than this threshold, it is implausible that the treated and control districts evolve similarly enough with respect to this outcome over time to warrant extrapolation. 
Based on detailed domain-knowledge, other values for the threshold may be reasonable.
For the third outcome, \emph{Agricultural Center}, our pretest $\phi = \indicator{ \estSpre \geq M }$ declares the extrapolation condition to be false; in this case, extrapolation is not warranted and so our analysis for this outcome comes to an end without any statement about the treatment effect.
For the first two outcomes, \emph{Educational and Cultural Program} and \emph{Tap Water}, our pretest declares that the extrapolation condition is true, so that we perform inference under the conditional extrapolation assumption and report our confidence interval $\ci$. 

Figure~\mainref{fig:malesky} compares our inference under the conditional extrapolation assumption with classical DID inference.
For the first two outcome measures, note that the proposed confidence intervals are wider than the conventional DID intervals.
This is because the proposed confidence intervals based on the conditional extrapolation assumption take into account both the magnitude of the severity of parallel trend violations in the pre-treatment period and the corresponding uncertainty in estimating this magnitude.
For the third outcome, we reject the extrapolation condition: based on DID, nothing can be said about the effect of treatment on this outcome.

\section{Sensitivity Analysis for the Conditional Extrapolation Assumption}
\label{app:sens}
In this section, we will develop sensitivity analysis for the conditional extrapolation assumption (Assumption \ref{a:extrapolation}).
Consider the following $\gamma$-inflated extrapolation assumption
\begin{equation}
\Spre \leq M \text{ , then } \Spost \leq \gamma\cdot \Spre,
\end{equation}
where $\gamma \geq 1$ is a sensitivity parameter. For a given $\gamma$, our confidence interval becomes 
\[
\hat C_{\level}^{(\gamma)} = \esttotaldd \pm \bigg \{\gamma \cdot \extrapbias \cdot \estSpre
        	+ \frac{ f_\gamma(\level, \estcov) }{ \sqrt{n}} \bigg \},
\]
which takes into account the larger possible worst-case bias due to $\gamma$, and a modified critical value $f_\gamma(\level, \estcov)$ accounting for the joint statistical fluctuations of $\esttotaldd$ and $\estSpre$. This critical value $f_\gamma(\level, \estcov)$ is the upper $\level$-quantile of the random variable $\psi_\gamma(Z)$, where 
\[
\psi_\gamma(x_1 \dots x_{T-1})  = \abs[\Big]{ \frac{1}{\Tpost} \sum_{t=T_{pre}}^{T-1} x_t }
+ \gamma \cdot \extrapbias \cdot \paren[\Big]{ \frac{1}{\Tpre-1} \sum_{t=1}^{T_{pre}-1} \abs{ x_t }^p }^{1/p}
\enspace,
\]
and $Z \sim \mathcal{N}(0,\estcov)$. That is,

\[
\Pr{
\psi_\gamma(Z) \geq f_\gamma(\level,\estcov)
}
=
\level,
\qquad
Z\sim \mathcal N(0,\estcov).
\]

It is straightforward to see that for any fixed $\gamma$, the confidence interval $\hat C_{\level}^{(\gamma)}$ enjoys the same conditional coverage guarantees of Theorem $\ref{theorem:cond-inf}$ under the $\gamma$-inflated conditional extrapolation assumption. The proofs follow exactly as for the original (un-inflated) confidence interval.

For a two-sided test of $H_0: \totalatt = 0$ at significance level $\level$, the result is significant under sensitivity value $\gamma$ if and only if $0 \notin \hat C_\level^{(\gamma)}$. 
Because both the bias-adjustment and the critical value $f_\gamma(\level, \estcov)$ are non-decreasing in $\gamma$, statistical significance is monotone in the sensitivity parameter. That is, if it is no longer significant for some $\gamma$, then it will not become significant for any larger $\gamma$.
We can thus define the \emph{overturning sensitivity value} at level $\level$ as
\[
\gamma_\level^{\text{over}} = \inf \bigg \{\gamma \geq 1: \lvert \esttotaldd \rvert \leq \gamma \extrapbias\estSpre + \frac{ f_\gamma(\level, \estcov) }{ \sqrt{n}} \bigg\},
\]
where for all $\gamma < \gamma_\level^{\text{over}}$ the result remains significant at level $\level$. If the result is already insignificant at $\gamma =1$, then $\gamma_\level^{\text{over}} = 1$.

This overturning value can be found numerically by solving the equation 
\[
\lvert \esttotaldd \rvert = \gamma\kappa\estSpre + \frac{ f_\gamma(\level, \estcov) }{ \sqrt{n}}.
\]
	
\end{document}